%% file: ms.tex
\documentclass[12pt]{article}  

\pdfoutput=1

\input{./pagelayoutWOmargins2.tex}

\usepackage{bm}
\usepackage{amsfonts}  
\usepackage{graphicx}
\usepackage{natbib}

\begin{document}

\def\spacingset#1{\renewcommand{\baselinestretch}%
{#1}\small\normalsize} \spacingset{1}

\title{\bf On the Hauck-Donner Effect in Wald Tests: Detection,
Tipping Points, and Parameter Space Characterization}
\author{Thomas~W.~Yee,
Department of Statistics, University of Auckland\\
(E-mail: \textit{t.yee@auckland.ac.nz})
}
  \maketitle

\spacingset{1.0}

\noindent
\textbf{Abstract} \qquad
The Wald test remains ubiquitous in statistical practice despite
shortcomings such as
its inaccuracy in small samples
and lack of invariance under reparameterization.
This paper develops on another but lesser-known shortcoming called the
Hauck--Donner effect (HDE) whereby a Wald
test statistic is not monotonely increasing as a function of
increasing distance between the parameter estimate and the null value.
Resulting in an upward biased $p$-value and loss of power,
the aberration
can lead to very damaging consequences such as in variable selection.
The HDE afflicts many types of regression models and corresponds to
estimates near the boundary of the parameter space.
This article presents several new results, and its main contributions are to
(i)~propose a very general test for detecting the HDE,
regardless of its underlying cause;
(ii)~fundamentally characterize the HDE by pairwise
ratios of Wald and Rao score and likelihood ratio test statistics
for 1-parameter distributions;
(iii)~show that the parameter space may be partitioned
into an interior encased by~5 HDE severity measures
(faint, weak, moderate, strong, extreme);
(iv)~prove that a necessary condition for the HDE in
a 2 by 2 table is a log odds ratio of at least~2;
(v)~give some practical guidelines about HDE-free
hypothesis testing.
Overall, practical post-fit tests can now be conducted
potentially to any model estimated by
iteratively reweighted least squares,
such as the generalized linear model (GLM) and Vector GLM (VGLM)
classes, the latter which encompasses many popular regression models.

\bigskip


\bigskip

\bigskip

\section{Introduction}

In classical likelihood theory three test statistics are used for
general hypothesis testing and inference.  They are the likelihood
ratio test (LRT), Rao's score (Lagrange multiplier) test, and the Wald
test.  It is well-known the LRT generally performs the best, and
that the Wald test suffers from shortcomings such as its lack of
invariance under reparameterization and inaccuracy in small samples.
Despite these, the Wald test is probably the most widespread test, as
output such as Table~\ref{tab:birds.waldtable} is ubiquitous in
statistical practice.

\begin{figure}[tt]
\begin{center}
\includegraphics[width=0.9\textwidth]{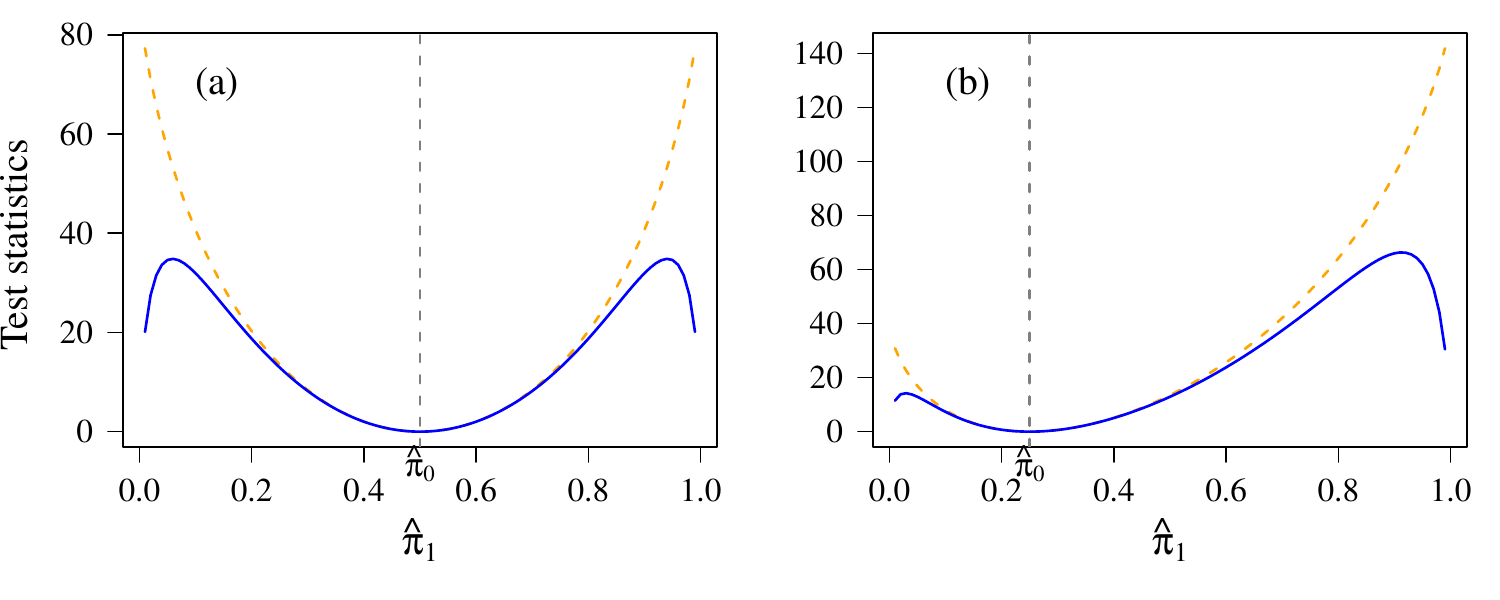}
\end{center}
\caption{
Wald and likelihood ratio test 
statistics plotted against~$\widehat{\pi}_1$, for:
(a)~$\widehat{\pi}_0=0.5$,
(b)~$\widehat{\pi}_0=0.25$ (vertical dashed lines).
These are $\calW_2$ (solid curves) and $\calW_L$ (dashed curves).
Here, $\widehat{\pi}_1 = 0.01$ to~$0.99$ in steps of~$0.01$
is discrete.
The data follows~\cite{hauc:donn:1977};
see also Table~\ref{tab:Cgelimo:hauc:donn:1977.dataset}(a).
\label{fig:Cgelimo:hauck.donner}
}
\end{figure}

A lesser known but no less pernicious problem of the
Wald test is
that it suffers potentially from the Hauck-Donner effect
\citep[HDE;][]{hauc:donn:1977,hauc:donn:1980} whereby the test
statistic fails to increase monotonically as a function of its
distance from the null value, e.g., as in
Figure~\ref{fig:Cgelimo:hauck.donner}.  Consequently, a truly large
effect contaminated by the HDE might be construed as being
nonsignificant.  The loss of power problem is aggravated by variable
selection procedures that are based on Wald's test.

In general likelihood theory,
much of statistical inference is
dichotomized into the cases where the true parameter value is either
in the interior of parameter space~$\bTheta$ or lies on its boundary.
This article is concerned with the interface of the two.
Informally, partition the closure
$\overline{\bTheta} = \bTheta^o \cup \partial \bTheta$
into its interior and boundary,
and further partition
$\bTheta^{o}= \bTheta^{oo} \cup \bTheta^{\mathrm{hde}}$
where~$\bTheta^{\mathrm{hde}}$ may be the empty set.
The usual regularity conditions fully operate
\textit{practically} in~$\bTheta^{oo}$
but start breaking down in~$\bTheta^{\mathrm{hde}}$
for want of sufficient Taylor series terms,
which we define as the subspace where an aberration
of the Wald statistic occurs.
Failure to distinguish between~$\bTheta^{oo}$
and~$\bTheta^{\mathrm{hde}}$ can result in incorrect inferences,
therefore special attention should be made to identify it if it
occurs.
This article sheds light on~$\bTheta^{\mathrm{hde}}$
(Figure~\ref{fig:Cgelimo:hdeff.paramspace}).

One reason for choosing and
defining~$\bTheta^{\mathrm{hde}}$ as such is the huge
popularity of Wald tests in
general regression modelling, including generalized linear models
\citep[GLMs;][]{neld:wedd:1972}.
They appear almost universally
as standard computer output in the
form of a~4-column matrix (called a {Wald table} in this article)
consisting of the point estimates, standard errors, Wald statistics,
and $p$-values, with optional embellishments of astericks and dots.

Despite four decades since it was first observed, there has been
relatively very little general
work characterizing the Hauck-Donner phenomenon, or even to detect it.
This work is an attempt to address these.
We use `general' because special cases such as $N\mu(1-\mu)>10$
for the (normal approximation to the)
binomial are very well-known special cases only.
An empirical approach is taken in this article and we
develop new methods
that can be used routinely on any GLM or potentially any model based
on a weighted crossproduct matrix of the form~$\bX^T \bW \bX$ ($=\bA$,
say) within an iteratively reweighted least squares (IRLS) algorithm.
Consequently the results are very general and widely applicable for
HDE detection.

\cite{vaet:1985} studied the HDE but mainly restricted his
investigation to the one-sample problem for one-parameter exponential
families.  His results were heavily
dependent on the nice mathematical properties of exponential families
and it was stated that these do not easily generalize to arbitrary
families of distribution functions.  He showed that, in general, the
Wald test statistic~$\calW$ should be used cautiously with discrete
probability models due to certain boundary problems, and that the test
may also be misleading in certain families of continuous
distributions.  He also advised that Wald's test should be used
cautiously in logistic regressions with many explanatory variables.
This caution has been repeated by others,
e.g., \cite{xing:etal:2012} in the context of
genome-wide case-control association studies for genetic screening
suggested the LRT as a better alternative.

It is stressed here, as does \cite{vaet:1985}, that it
is the behaviour of the Wald statistic for~$\widehat{\beta}_s$,
for fixed sample size~$n$,
as the maximum likelihood estimate
(MLE) moves away from the null value that is
of concern and not the distributional properties of~$\calW$
as~$n \rightarrow \infty$.
This article concerns simple null hypotheses of the form
$H_{0s}: \beta_s = \beta_{s0}$ for some known
prespecified~$\beta_{s0}$ that is usually taken to be~0;
the signed square root of the Wald statistic for the~$s$th coefficient
is~$\widetilde{\calW}_s=
(\widehat{\beta}_s-\beta_{s0})/\SE(\widehat{\beta}_s)$
where the Wald statistic~$\calW_s=\widetilde{\calW}_s^2$
is asymptotically~$\chi_1^2$ under~$H_0$.
Note that while the denominator of~$\widetilde{\calW}_s$ is sometimes
evaluated at~$\beta_{s0}$
(in which case the HDE will be absent), the
vast majority of software such as \texttt{glm()} in \RR{}
evaluate the standard error SE at the MLE so that the HDE is
an ever-present threat.

Notationally,
let the~$(j,s)$ element of the inverse of a
matrix~\bA{} having~$(j,s)$ element~$a_{js}$ be denoted by~$a^{js}$,
and $\bA \succ \bO$
and~$\bA \succeq \bO$ indicates that the symmetric
matrix~\bA{} is
positive-definite and
positive semi-definite
respectively.
The Hadamard (element-by-element)
and Kronecker products of two
matrices~\bA{} and~\bB{}
are denoted
by~$\bA \circ \bB = [(a_{js} \cdot b_{js})]$
and~$\bA \otimes \bB = [(a_{js} \cdot \bB)]$
respectively.
If \btheta{} and $\boldeta$ are $M$-vectors
then $\partial \btheta / \partial \boldeta^T$ is
an order-$M$ matrix whereas
$\partial \btheta / \partial \boldeta$ is
an $M$-vector with elements~$\partial \theta_j / \partial \eta_j$.
We also let~$\bie_{j}$ be a vector of~0s except for a~1 in
the~$j$th position, whose dimension is obvious,
and the indicator function~$I_{[i=j]}$
equals~1 if~$i=j$ and~0 otherwise.
We use~$\calW_L= 2(\ell - \ell_0)$ to denote the LRT statistic.

An outline of the paper is as follows.
As a concrete example,
we consider the original data set of
\cite{hauc:donn:1977} in
Section~\ref{sec:Cgelimo:hauc:donn:1977.dataset}
(as well as citing more instances of the HDE by others
for additional motivation)
before describing elements of a class of
models called VGLMs---the detection
test specifically applies to this (large) class.
The general test is described for VGLMs in
Section~\ref{sec:Cgelimo:hdeff.M.gt.1},
and supporting asymptotic results are given in
Section~\ref{sec:Cgelimo:hdeff.inference}.
Section~\ref{sec:Cgelimo:hdeff.refinements} proposes
two important refinements:
finite-difference approximations for derivatives to make
HDE detection applicable to all VGLMs, and
categorizing the HDE into~4 severity measures that form
a partition of~\bTheta.
Some numerical examples are given in
Section~\ref{sec:Cgelimo:hdeff.eg}, and
Section~\ref{sec:Cgelimo:hdeff.summary} discusses
computational details and provides some practical guidelines.
The paper concludes with a discussion of the overall findings.
The methodology here is implemented in the \RR{} package
\pkg{VGAM}~1.0-6 or later (available on CRAN) for 100+ models.
Some extensions for detecting the HDE in related contexts
are given in Appendices A2--A5.

\section{The Hauck \& Donner Data Set}
\label{sec:Cgelimo:hauc:donn:1977.dataset}

\begin{figure}[pp]
\begin{center}
\includegraphics[width=\textwidth]{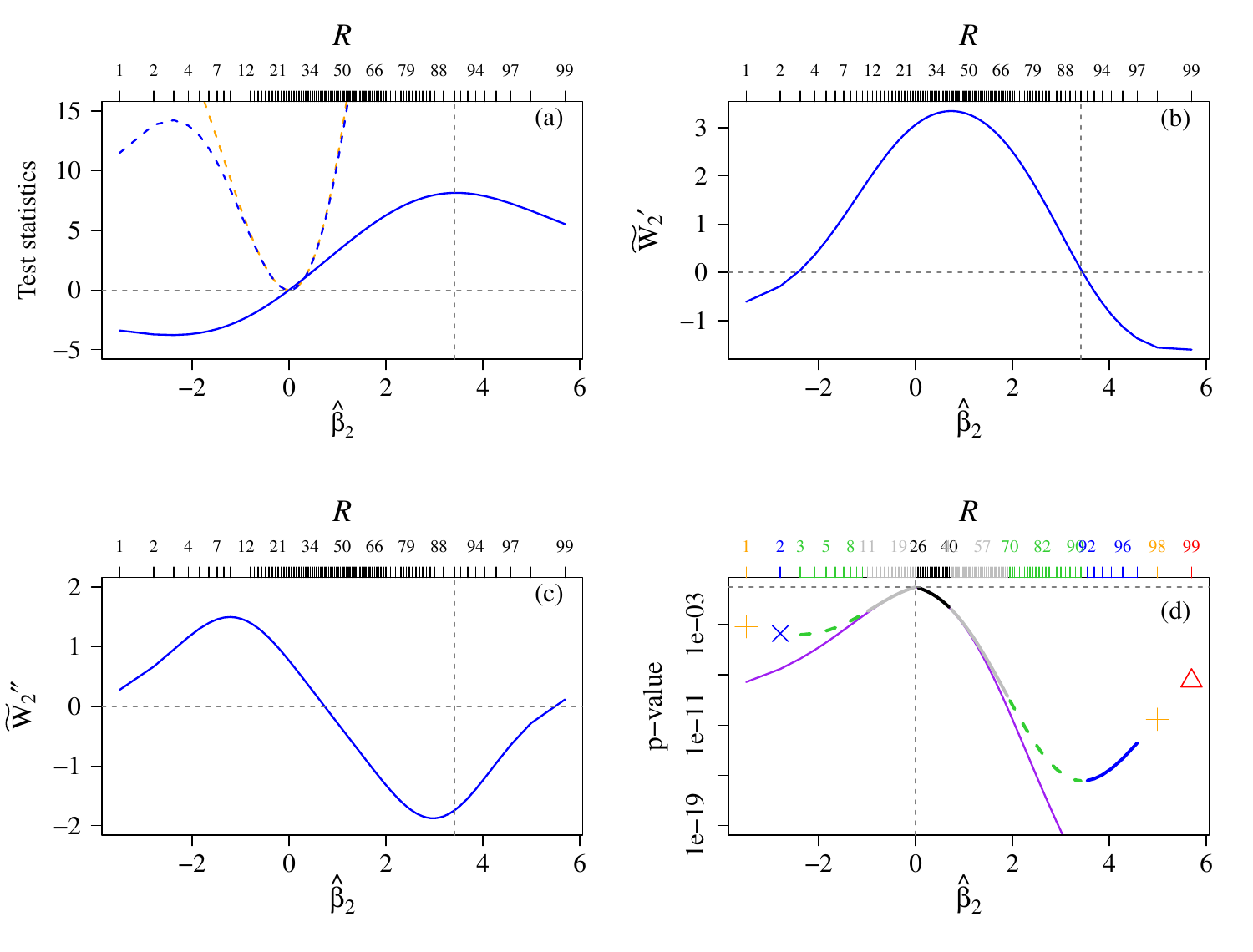}
\end{center}
\caption{
The Hauck \& Donner data: how the (signed root) Wald test
statistic~$\widetilde{\calW}_2$
varies as a function of~$\widehat{\beta}_2$ (bottom axis)
and~$R$ (top axis).
This corresponds to Figure~\ref{fig:Cgelimo:hauck.donner}(b).
The vertical dashed lines correspond to~$R=91$.
(a)~The original~$\widetilde{\calW}_2$ is the solid curve,
$\calW_2$ is dashed curve and is~$\chi_1^2$ under~$H_0$,
and the quadratic-shaped dashed curve is the LRT statistic~$\calW_L$.
(b)~First derivative~$\widetilde{\calW}_2'(\widehat{\beta}_2)$.
(c)~Second derivative~$\widetilde{\calW}_2''(\widehat{\beta}_2)$.
(d)~LRT (lower purple curve)
and Wald test p-values with the latter
color-coded as in Fig.~\ref{fig:Cgelimo:hdonner.severity};
a log-scale is used and the horizontal reference line is at p-value $=1$.
\label{fig:Cgelimo:hdonner.sims}
}
\end{figure}

\cite{hauc:donn:1977} apply logistic regression to a
$2 \times 2$ table of counts
to illustrate the HDE. Because of its simplicity a more complete analysis
of the behaviour of~$\widetilde{\calW}_2$ is permitted here.
We will generalize the data slightly to afford
a little more flexibility
(Table~\ref{tab:Cgelimo:hauc:donn:1977.dataset}(a)).
The actual data has~$N=100$,
$n=2N$,
$R_{0}=25$,
and we will generally
keep these fixed but vary~$R\in \{1,\ldots,N-1\}$.
The sample proportions are~$\widehat{\pi}_0 = R_{0}/N = 0.25$ at~$x_2=0$,
and~$\widehat{\pi}_1 = R/N$ at~$x_2=1$.
Fit
\begin{eqnarray}
  \label{eq:Cgelimo:hauc:donn:1977.model}
\logit \, \mu_i ~=~
  \beta_1 + \beta_2\, x_{i2},  ~~~
  i=1,\ldots,n, ~~~ \mathrm{independently},
\end{eqnarray}
so that
$\widehat{\beta}_1 = \logit \, \widehat{\pi}_0 \approx -1.099$  
and
$\widehat{\beta}_2 = \log \widehat{\psi}$ is the estimated log odds ratio.
The HDE becomes pronounced once past a certain threshold
as~$R\rightarrow 0$ or~$R\rightarrow N$,
corresponding to sparsity in one off-diagonal cell.
In Figure~\ref{fig:Cgelimo:hdonner.sims}
the larger threshold corresponds to~$R\approx 91$.
For such count tables the well-known formula
\begin{eqnarray}
\SE(\log \widehat{\psi}) ~\approx~
\sqrt{
  \frac{1}{N-R_0} +
  \frac{1}{R_0} +
  \frac{1}{N-R} +
  \frac{1}{R}
}
\label{eq:Cgelimo:SE.log.oration}
\end{eqnarray}
is routinely used.
For handling low counts,
there are several popular \textit{pre-fit} improvisions,
e.g., adding~0.5 to each
cell count in the SE calculation give a bias-corrected log odds ratio,
and applying the Mantel-Haenszel method.
However, in the context of routine logistic regressions, it is far
more convenient for some \textit{post-fit}
test for HDE to be applied to a fitted
model---if the HDE is absent then we can conclude that the counts are
sufficiently large that no adverse behaviour on the SE was seen.

In this particular example the HDE can be explained by a situation
already well-known to practitioners.
This is not the case in general because the HDE has been
observed in other regression models by various authors since.
Some examples include
\cite{stor:wach:bres:1983} in conditional logistic regression
with matched and stratified samples,
\cite{vaet:1985} in one-sample problems for
one-parameter exponential families and GLMs,
\cite{nels:savi:1990} in Tobit and nonlinear regression models,
\cite{fear:beni:gail:1996} in a balanced 1-way random effects
ANOVA design,
\citet[p.60]{ther:gram:2000} in Cox proportional hazards models,
and
\cite{kosm:2014} in cumulative link models.
In general the Wald test can be expected to be valid only if a
normal likelihood can be used to approximate the profile likelihood
for the parameter well \citep{meek:esco:1995}
and the observed value of the sufficient statistic is away
from~$\partial \bTheta$.

Regardless of its underlying cause, it would be very useful to have a
procedure for detecting the HDE that could be routinely applied to a
given regression model.  To this end,
Section~\ref{sec:Cgelimo:hdeff.M.gt.1} proposes a method for this, and
the method applies very generally to models estimated by IRLS.  This
general purpose algorithm has been used to fit many popular regression
models, and most notable is the GLM class, of which a multivariate
extension called vector GLMs \citep[VGLMs;][]{yee:2015} have been
proposed.

Continuing with this example,
when the full model is fitted by IRLS, then
\begin{eqnarray}
    \label{eq:Cgelimo:xtwx.inv.binomial}
\ \ ~~~
\bA^{-1} ~=~
\frac{1}{N}
\left(
\begin{array}{cc}
  \widehat{\pi}_0^{-1} (1-\widehat{\pi}_0)^{-1} &
 -\widehat{\pi}_0^{-1} (1-\widehat{\pi}_0)^{-1} \\
 -\widehat{\pi}_0^{-1} (1-\widehat{\pi}_0)^{-1} &
~~\widehat{\pi}_0^{-1} (1-\widehat{\pi}_0)^{-1} +
  \widehat{\pi}_1^{-1} (1-\widehat{\pi}_1)^{-1}~
\end{array}
\right).
\end{eqnarray}
Consequently,
(\ref{eq:Cgelimo:SE.log.oration}) is obtained,
as is
\begin{eqnarray}
\frac{\partial \widetilde{\calW}_2}{\partial \widehat{\beta}_2}
&~=~&
\frac{1}{\SE(\widehat{\beta}_2)}
\left\{
1 -
\frac{ \widehat{\beta}_2 \cdot \left( 2 \widehat{\pi}_1 - 1 \right) }{
  N \, \widehat{\pi}_1  (1 - \widehat{\pi}_1) \; \SE^2(\widehat{\beta}_2)}
\right\},
  \label{eq:Cgelimo:hauc:donn:1977.dW.dbeta2}
\end{eqnarray}
which is plotted in Figure~\ref{fig:Cgelimo:hdonner.sims}(b).
It appears impossible to write a closed form expression for~$R$
upon setting~(\ref{eq:Cgelimo:hauc:donn:1977.dW.dbeta2}) to~0;
this indicates that it is impractical
trying to determine beforehand the threshold
of how low the counts can be before observing the HDE---and
likewise for more complicated situations.
This supports the view that a detection test
post-fit is more practical than trying to avoid it pre-fit.

Incidentally,
the LRT cannot exhibit the HDE
\begin{eqnarray*}
\frac{\partial^2 \calW_L}{\partial R^2}
&~=~&
2
\left(
\frac{1}{R} - \frac{1}{R + R_0}
\right) +
2
\left(
\frac{1}{N-R} - \frac{1}{(N-R) + N - R_0}
\right) ~>~ 0
\end{eqnarray*}
for all~$R \in \{1,\ldots,N-1\}$,
provided that~$R_0>0$,
therefore~$\calW_L$ is convex in~$\widehat{\beta}_2$.
It will also be seen that the Rao score (Lagrange multiplier) test is
also immune to the HDE as it does not depend on the third derivatives
of the log-likelihood.

\renewcommand{\arraystretch}{1.4}
\begin{table}[tt]
\caption{
(a)~The Hauck \& Donner data with a slight generalization for added
flexibility; counts in a $2\times 2$ table
(Usually~$N=100$, $R_0=25$; and~$R>91$ exhibits the HDE).
(b)~A generalization of~(a) allowing for disproportional sampling.
The quantities
are~$f_0 = N_0 / N_1^*$,
and~$N_1^*=c^* N_1$,
for some multiplier~$c^* \geq 1$.
\label{tab:Cgelimo:hauc:donn:1977.dataset}
}
\centering
 \begin{tabular}{lccrlccl}
~~~(a) & $y=0$ & $y=1$
& ~~~~~~~~~~~ &
~~~(b) & $y=0$ & $y=1$ & \\
\cline{2-3}
\cline{6-7}
\multicolumn{1}{l|}{$x_{2}=0$~~} &
\multicolumn{1}{c|}{$N-R_0$} &
\multicolumn{1}{c|}{~~~$R_0$~~~}
& ~~~~~~~~ &
\multicolumn{1}{l|}{$x_{2}=0$~~} &
\multicolumn{1}{c|}{$N_0-R_0$} &
\multicolumn{1}{c|}{~~~$R_0$~~~}
& ~$N_0$
\\
\cline{2-3}
\cline{6-7}
\multicolumn{1}{l|}{$x_{2}=1$~~} &
\multicolumn{1}{c|}{$N-R$} &
\multicolumn{1}{c|}{~~~$R$~~~}
& ~~~~~~~~ &
\multicolumn{1}{l|}{$x_{2}=1$~~} &
\multicolumn{1}{c|}{$c^* (N_1 - R)$} &
\multicolumn{1}{c|}{~~~$c^* R$~~~}
& ~$N_1^*$
   \\
\cline{2-3}
\cline{6-7}
\end{tabular}
\end{table}
\renewcommand{\arraystretch}{1.0}

\begin{figure}[tt]
\begin{center}
\includegraphics[width=\textwidth]{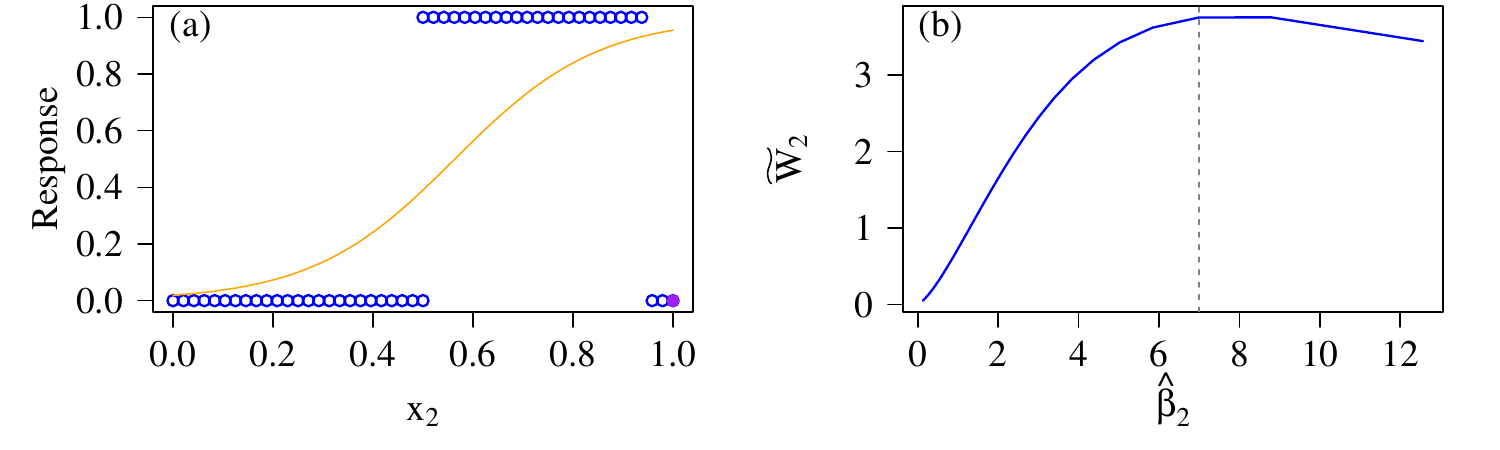}
\end{center}
\caption{
HDE arising from an approximate quasi-complete separable data set
with $n=50$.
The~$x_{i2}$ are equally spaced on~[0,\ 1] plus one more
value at~$\frac12$.
(a)~Some of the~$x_{i2}$ values beyond~$\frac12$
have had their~$y_i=0$ values
replaced by~$y_i=1$; there are
$3 - 1$ remaining at this stage---and the
solid rightmost point remains unchanged.
The curve are the fitted values of the logistic regression.
(b)~Signed Wald statistics~$\widetilde{\calW}_2(\widehat{\beta}_2)$
over the data sets.
The vertical dashed line corresponds to the data in~(a).
\label{fig:Cgelimo:hdonner.qsep}
}
\end{figure}

Somewhat similar to the Hauck \& Donner data set,
it is noted that the HDE can arise as a result of data exhibiting near
overlap, quasi-complete separation or complete separation
\citep[see, e.g.,][]{albe:ande:1984,lesa:albe:1989}.
For example,
starting off with a data set
comprising
$(x_{i2} = (i-1)/(2N-2),\ y_i=0)$,
$i=1,\ldots,2N-1$,
plus~$(x_{N2}=\frac12,\ y_i=1)$,
we replace each successive~$y_i=0$ on the RHS
of~$x_{i2}>\frac12$ by~$y_i=1$
(Figure~\ref{fig:Cgelimo:hdonner.qsep})
except for the very rightmost.
As the number of~$y_i=1$ increases
the logistic regression~(\ref{eq:Cgelimo:hauc:donn:1977.model})
fitted to these data has~$\widehat{\beta}_2$ increasing
and the HDE will become present eventually.
Figure~\ref{fig:Cgelimo:hdonner.qsep}(b) shows the
Wald statistic as a function of~$\widehat{\beta}_2$, and the
HDE is evident at the RHS.
Data separation is likely in Big Data situations
when extraneous covariates are included in regression
models of high dimensionality.

\subsection{VGLMs}

The detection test specifically applies to this (large) class,
therefore we briefly summarize them.
They can loosely be thought of as multivariate GLMs applied
to parameters~$\theta_j$
but extending far beyond the exponential family;
more details can be found in \cite{yee:2015}.
The data is~$(\bix_i,\ \biy_i)$,
$i=1,\ldots,n$, independently,
with response~$\biy_i$
and usually with intercept~$x_{i1}=1$.
The~$j$th~linear predictor is
\begin{equation}
g_j(\theta_j) ~=~
\eta_j ~=~ \bbeta_j^{T} \bix ~=~
\sum_{k=1}^d \; \beta_{(j)k} \, x_{k}, ~~~ j=1, \ldots, M,
\label{gammod2}
\end{equation}
for some \textrm{parameter link function}~$g_j$
satisfying the usual properties
(strict monotonicity and twice-differentiable).
If~$M>1$ then linear constraints between
the regression coefficients are accommodated, as
\begin{eqnarray}
\boldeta(\bix_i)
  & = &
\left(
\begin{array}{c}
\eta_1(\bix_i) \\
\vdots\\
\eta_M(\bix_i)
\end{array}
\right) ~=~
\sum_{k=1}^d \, \bbeta_{(k)} \, x_{ik} ~=~
\sum_{k=1}^d \, \bH_k \, \bbeta_{(k)}^{*} \, x_{ik},
\label{eq:constraints.VGLM}
\end{eqnarray}
for known {constraint matrices}~$\bH_k$
of full column-rank (i.e., rank~$\mathcal{R}_k=$ \rcodett{ncol}($\bH_k$)),
and~$\bbeta_{(k)}^{*}$
is a possibly reduced set of regression coefficients
to be estimated.
While trivial constraints are denoted by~$\bH_k=\bI_M$, other
common examples include parallelism ($\bH_k=\bone_M$), exchangeability,
and intercept-only parameters~$\eta_j=\beta_{(j)1}^*$.
The overall `large' model matrix is~$\bX_{\stVLM}$, which
is~$\bX_{\stLM} \otimes \bI_M$ with trivial constraints,
while~$\bX_{\stLM}=[(x_{ik})]$ is the `smaller'~$n \times d$ model
matrix associated with a $M=1$ model.

Some models have $\eta_j$-specific explanatory variables,
such as a time-varying covariate,
then~(\ref{eq:constraints.VGLM}) extends to
\begin{eqnarray}
\boldeta_i
&~=~&
\bio_i +
\sum_{k=1}^{d} \, \diag(x_{ik1},\ldots,x_{ikM}) \,
\bH_k \, \bbeta_{(k)}^{*}
\label{eq:Cvglimo:xij.vector.diag}
\end{eqnarray}
with provision for offsets~$\bio_i$.
The results of this paper apply most generally to this case.

The~$\bW^{(a)}=(\bW_1^{(a)},\ldots,\bW_n^{(a)})$ are the working weight
matrices, comprising $\bW_i^{(a)} = \mbox{}$
$-E[\partial^2 \ell_i / (\partial \boldeta_i\,\partial \boldeta_i^T)]$
at iteration~$a$,
for each log-likelihood component~$\ell_i$.
Here, $\ell=\sum_{i=1}^n \ell_i$ is the log-likelihood, and
Fisher scoring is adopted as opposed to Newton-Raphson.
Usually the individual expected information matrices (EIMs)
are closely related to the working weight matrices~$\bW_i$ as
$\EIMsub{i} \circ
  \left[
  ({\partial \btheta  }/{\partial \boldeta_i  })
  ({\partial \btheta^T}/{\partial \boldeta_i^T})
  \right]$,
i.e.,
\begin{eqnarray}
\label{eq:hdeff:eim.ab}
(\bW_i)_{uv}
&~=~&
-E\left[
  \frac{ \partial^2 \ell_i}{\partial \eta_u\,\partial \eta_v}
  \right] ~=~
 E\left[
  \frac{-\partial^2 \ell_i}{\partial \theta_u\,\partial \theta_v}
  \right]\;
  \frac{\partial \theta_u}{\partial \eta_u} \;
  \frac{\partial \theta_v}{\partial \eta_v} \\
&~=~& \nonumber
(\EIMsub{i})_{uv} \cdot
\left[ \; g'_u(\theta_u) \; \, g'_v(\theta_v) \right]^{-1},~~~
\mathrm{say},
\end{eqnarray}
for~$u,v\in\{1,\ldots,M\}$.
In particular, (\ref{eq:hdeff:eim.ab}) holds for 1-parameter
link functions~$g_u$.

For VGLMs
the estimated variance-covariance matrix is
\begin{eqnarray}
\widehat{\Var}
\left(
\widehat{\bbeta^{*}}
\right) &~=~&
\left(
\bX_{\stVLM}^T \;
\bW^{(a)} \;
\bX_{\stVLM}
\right)^{-1},
\label{eq:vglimo:vcov.coef.copy2}
\end{eqnarray}
evaluated at the final iteration,
where~$\bbeta^{*}=(\bbeta_{(1)}^{*T},\ldots,\bbeta_{(d)}^{*T})^T$
are all the regression coefficients to be estimated.
The iteration number~$a$ will be suppressed henceforth.
One reason for the widespread use of the Wald test is their
computationally convenience: the estimated variance-covariance
matrix~(\ref{eq:vglimo:vcov.coef.copy2}) is a by-product of the IRLS
algorithm, and importantly it is evaluated at the
MLE~$\widehat{\bbeta^{*}}$.

\section{HDE Detection}
\label{sec:Cgelimo:hdeff.M.gt.1}

The overall model matrix has form
$\bX_{\stVLM} =\left( \bX_1^T,\ldots,\bX_n^T \right)^T$
so that~$\bA = \sum_{i=1}^n \; \bX_i^T \, \bW_i \, \bX_i$
and
\begin{eqnarray}
  \label{eq:hdeff.xtwx.deriv1}
\frac{\partial^{\nu} \bA}
{\partial \, \widehat{\beta}_{(r)t}^{*\,\nu}} ~=~
\sum_{i=1}^{n} \;
\bX_{i}^T \;
\frac{\partial^{\nu} \, \bW_{i}}
{\partial \, \widehat{\beta}_{(r)t}^{*\,\nu}} \;
\bX_{i}
\end{eqnarray}
for~$\nu=1,2,\ldots$,
and~$r\in\{1,\ldots,\mathcal{R}_s\}$
and~$t\in\{1,\ldots,d\}$.
For simplicity map the coefficients
to~$(\widehat{\beta}_1,\widehat{\beta}_2,\ldots)$
so that for the $s$th coefficient
$0<a^{ss}$ because \bA{} is positive-definite.
Then~$(a^{ss})'$ can be computed by
\begin{eqnarray}
  \label{eq:deriv.matrix}
  \frac{\partial \bA^{-1}}{\partial \widehat{\beta}_s} ~=~
  -\bA^{-1} \;
  \frac{\partial \bA}{\partial \widehat{\beta}_s} \;
   \bA^{-1}.
\end{eqnarray}
Of central interest
for testing~$H_0: \beta_s = \beta_{s0}$
versus~$H_1: \beta_s \neq \beta_{s0}$ is
\begin{eqnarray}
\frac{ \partial \widetilde{\calW}_s}{\partial \widehat{\beta}_s}
&~=~&
\frac{ \partial  }{\partial \widehat{\beta}_s}
\frac{ \widehat{\beta}_s - \beta_{s0} }{\sqrt{ a^{ss}}}
\ =~
\frac{1}{\sqrt{ a^{ss}}}
\left[ 1 -
\frac{\widehat{\beta}_s - \beta_{s0} }{2}\,
\frac{(a^{ss})'}{a^{ss}}
\right].
\label{eq:Cgelimo:waldstat.deriv1}
\end{eqnarray}
This equation furnishes a first-derivative detection test:
the HDE is evident for~$\widehat{\beta}_s$
if~(\ref{eq:Cgelimo:waldstat.deriv1})
is negative-valued.
Consequently that coefficient's SE and $p$-value should be
flagged as unreliably biased upwards.
Only a quadratic form needs to be computed in~(\ref{eq:deriv.matrix})
for each~$a^{ss}$.

With provision to handle constraint matrices
and~$\bix_{ij}$ as in~(\ref{eq:Cvglimo:xij.vector.diag}),
\begin{eqnarray}
  \frac{\partial \bW_i}{\partial \widehat{\beta}_{(r)s}^{*}}
  &~=~&
  \label{eq:hdeff:dW.dbetas}
  \sum_{j=1}^M \;
  \frac{\partial \bW_i}{\partial \theta_j}\,
  \frac{\partial \theta_j}{\partial \eta_j}\,
  \frac{\partial \eta_j}{\partial \widehat{\beta}_{(r)s}^{*}} \\
  &~=~&
  \nonumber
  \sum_{j=1}^M \;
\left[
  \frac{\partial \, \EIMsub{i}}{\partial \theta_j} \circ
  \left(
  \frac{\partial \btheta  }{\partial \boldeta  } \,
  \frac{\partial \btheta^T}{\partial \boldeta^T}
  \right) +
  \mbox{} \right. \\ && \left. \nonumber
  \frac{\partial^2 \theta_j}{\partial \eta_j^2} \,
  \frac{\partial \eta_j}{\partial \theta_j}
  \cdot
  \EIMsub{i} \circ
  \left(
  \frac{\partial \btheta}{\partial \boldeta} \,
  \bie_j^T
  +
  \bie_j \,
  \frac{\partial \btheta^T}{\partial \boldeta^T}
\right)
\right]
  \frac{\partial \theta_j}{\partial \eta_j}
\;
  (\bH_s)_{jr} \;
  x_{isj}
\end{eqnarray}
because the working weight matrices have the simple
outer product form~(\ref{eq:hdeff:eim.ab}).

\subsection{Some Remarks and Properties}  

Several remarks are in order at this stage,
which mainly pertain to models with~$M=1$ parameter.

\begin{figure}[tt]
\begin{center}
\includegraphics[width=0.9\textwidth]{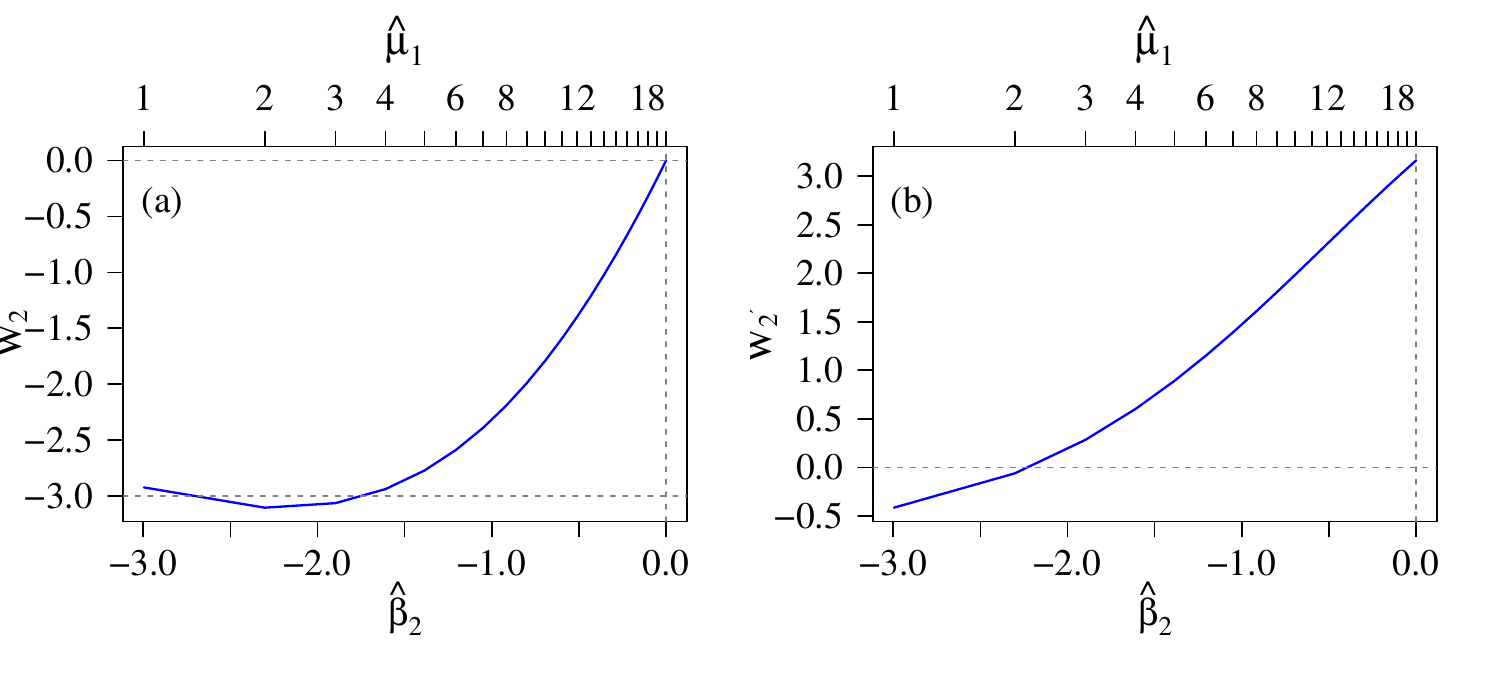}
\end{center}
\caption{
Simple Poisson regression example:
how the Wald test statistic~$\widetilde{\calW}_2$
varies as a function of~$\widehat{\beta}_2$ (bottom axis)
and~$\widehat{\mu}_1$ (top axis).
(a)~The original~$\widetilde{\calW}_2$ is the solid curve.
Dashed reference lines through the origin and at $-3$
have been added.
(b)~First derivative~$\widetilde{W}'(\widehat{\beta}_2)$.
\label{fig:Cgelimo:hdeff.eg.poissonff}
}
\end{figure}

\begin{enumerate}

\item
From~(\ref{eq:Cgelimo:waldstat.deriv1})
the Wald test statistic is aberrant if and only if
\begin{eqnarray}
\frac12 \left( \widehat{\beta}_s - \beta_{s0} \right) \,
\frac{d \, \log a^{ss}}{d \, \widehat{\beta}_s} - 1  ~>~ 0.
\label{eq:hdeff:aberrant.condit}
\end{eqnarray}
For 1-parameter models this criteria
simplifies to an interesting
expression involving the ratio of the Wald and LRT statistics
being less than the constant~$3/5$
(Section~\ref{sec:Cgelimo:hdeff.lrt.1parameter}).

\item
  The following are some sufficient conditions for~$\widetilde{\calW}_s$
  to be monotonic as~$\widehat{\beta}_s \rightarrow \infty$.
\begin{enumerate}

\item[(i)]
If
\begin{eqnarray}
  \label{eq:posdef.deriv.matrix}
  \frac{\partial \bA}{\partial \widehat{\beta}_s} ~\succ~ \bO.
\end{eqnarray}
The result follows from
a property of positive-definite matrices
\citep[e.g.,\ Eq.~(10.46)\ of][]{sebe:2008}
and
(\ref{eq:hdeff.xtwx.deriv1})--(\ref{eq:Cgelimo:waldstat.deriv1}).
One really needs~$(a^{ss})' < 0$ in~(\ref{eq:Cgelimo:waldstat.deriv1})
and this is provided if the inner matrix in~(\ref{eq:deriv.matrix}) is
positive-definite.  For~$M=1$ models, (\ref{eq:posdef.deriv.matrix})
entails that~$\partial^{} w_i / \partial \widehat{\beta}_s >0$
in~(\ref{eq:hdeff.xtwx.deriv1}).

\item[(ii)]
If~$(a^{ss})' = 0$.
For example, the full-likelihood LM
\begin{eqnarray}
  \label{eq:uninormal}
\eta_1=\mu,   ~~~~
\eta_2=\log\,\sigma, ~~~~
Y \sim N(\mu,\ \sigma^2),
\end{eqnarray}
whose EIM is diagonal so that each parameter can be treated separately,
the regression coefficients corresponding to~$\mu$ do not
suffer from the HDE because
the~(1,\ 1) element of the EIM
is not a function of~$\mu$.
For~$\eta_2$
the choice of link function \textit{does} matter,
e.g.,
using an identity link
then
${\partial w_i}/{\partial \widehat{\beta}_s}
= {-4\, x_{is}} \, {\sigma^{-3}}$,
so that if~$x_{is} > 0$ then
$\partial^{\nu} \bA / \partial \widehat{\beta}_s^{\nu} \prec \bO$
and
$\partial \bA^{-1} / \partial \widehat{\beta}_s \succ \bO$
so that~$(a^{ss})' > 0$;
thus ${\partial \widetilde{\calW}_s} / {\partial \widehat{\beta}_s} < 0$
if~$\widehat{\beta}_s \gg 0$.

\end{enumerate}

\item
For 1-parameter VGLMs, (\ref{eq:hdeff:dW.dbetas})
simplifies to
\begin{eqnarray}
  \nonumber
  \frac{\partial w_i}{\partial \widehat{\beta}_s}
  &~=~&
  \frac{\partial w_i}{\partial \theta}
  \frac{\partial \theta}{\partial \eta}
  \frac{\partial \eta}{\partial \widehat{\beta}_s}
  ~=~
  \label{eq:hdeff:dw.dbetas}
\left[
  \frac{\partial \, \EIMsub{i}}{\partial \theta} \;
  \left(
  \frac{\partial \theta}{\partial \eta}
  \right)^2 +
  2\; \EIMsub{i} \;
  \frac{\partial^2 \theta}{\partial \eta^2}
\right]
  \frac{\partial \theta}{\partial \eta}
\;
  x_{is} \; .
\end{eqnarray}
This makes allowance for a variety of link functions,
e.g., probit and complementary log-log links for binary regression.

\item
For standard logistic regression
$\partial^{} w_i / \partial \widehat{\beta}_s =
(1-2\mu_i) \, \mu_i (1-\mu_i) \, x_{is}$,
which is an odd function about~$\mu_i=0.5$.
Thus the model becomes more susceptible to the HDE
as~$\mu_i$ approaches a boundary
(observed in Figure~\ref{fig:Cgelimo:hdonner.sims}
as~$|\widehat{\beta}_2|$ becomes large.)
It is shown
that~$|\widehat{\beta}| > 2$
is a necessary condition for the HDE
in Section~\ref{sec:Cgelimo:hdeff.dispro}.
In fact, if~$\pi_0 = \frac12$ then
it is shown
in Appendix~A6
that approximately~$|\widehat{\beta}_2| > 2.40$ is needed
in order for the HDE to occur,
e.g., this corresponds to an odds ratio of about~$11.0$ or higher.

\label{item:Cgelimo:cumulants.binomialff}

\item
For the standard Poisson regression model,
applying a similar derivation as logistic regression
to a data set comprising~$N$ points at~$(x=0,\ y=\mu_0^{})$
and~$N$ points at~$(x=1,\ y=\mu_1^{})$,
yields
\begin{eqnarray}
  \label{eq:Cgelimo:poissonff.dW2.dbeta2}
\frac{\partial \widetilde{\calW}_2}{\partial \widehat{\beta}_2}
&~=~&
\sqrt{\frac{N \, \widehat{\mu}_0 \, \widehat{\mu}_1}
           {\widehat{\mu}_0 + \widehat{\mu}_1}}
\left\{
1 +
\widehat{\beta}_2 \;
\frac{\widehat{\mu}_0}{\widehat{\mu}_0 + \widehat{\mu}_1}
\right\}.
\end{eqnarray}
Thus conducive conditions for the HDE are when~$\widehat{\beta}_2 \ll 0$
and~$\widehat{\mu}_0 \gg \widehat{\mu}_1$.
Figure~\ref{fig:Cgelimo:hdeff.eg.poissonff} shows this
for~$N=1$, $\mu_0^{} = 20$
and~$\mu_1^{}$ taking on successive values in~$\{1,\ldots,20\}$.
It can be seen that if~$H_0$ were rejected when~$|\widetilde{\calW}_2|>3$
then it would do so only for~$\mu_1^{}=2$ or~3 but not~1.

\item
Given that~$\widetilde{\calW}_s'$ is computable, it can
be of interest to determine the rate at which
the $p$-value decreases as a function
of the effect size.
Assuming that $p_s \approx 2 \, \Phi(-|\widetilde{\calW}_s|)$
for a two-sided alternative, called~$p_s$ say,
then
\begin{eqnarray}
\label{eq:hdeff:pvalue.deriv1}
\frac{\partial p_s}{\partial \widehat{\beta}_s}
&~=~&
-2 \, \phi\left( \widetilde{\calW}_s \right) \,
\sgn(\widetilde{\calW}_s) \,
  \frac{\partial \widetilde{\calW}_s}{\partial \widehat{\beta}_s}.
\end{eqnarray}
An application might be in designing simple experiments
to determine how the $p$-value will decrease
given an increasing treatment effect.

\item
The basic
technique~(\ref{eq:hdeff.xtwx.deriv1})--(\ref{eq:hdeff:dW.dbetas})
carries naturally over to a variety of setting
such as sandwich estimators, multiple tests and profile likelihoods
(Appendices~A2--A5).

\item
If~$\theta_u$ and~$\theta_v$ are orthogonal
then the~$(u,\;v)$ element of the information matrix remains~0
upon differentiation, therefore agrees intuitively that the Wald statistic
for~$\widehat{\beta}_u$ is minimally affected
by~$\widehat{\beta}_v$.

\item
Second derivatives for the Wald statistic follow as before;
these can be used,
for example,
to determine whether~$\calW_s$ is a convex function,
i.e.,
is a particular model impervious to the HDE?
Use
\begin{eqnarray}
\frac{ \partial^2 \widetilde{\calW}_s}{\partial \widehat{\beta}_s^2}
&~=~&
\frac{1}{(a^{ss})^{3/2}}
\left[
-(a^{ss})' +
\frac{\widehat{\beta}_s - \beta_{s0}}{2}
\left\{
\frac32 \,
\frac{[(a^{ss})']^2}{a^{ss}} - (a^{ss})''
\right\}
\right]
\label{eq:Cgelimo:waldstat.deriv2}
\end{eqnarray}
and
\begin{eqnarray}
  \frac{\partial^2 \bA^{-1}}{\partial \widehat{\beta}_s^2}
  &~=~&
  \label{eq:deriv2.matrix.1param}
  \bA^{-1}
  \left[ \,
  2 \;
  \frac{\partial \bA^{}}{\partial \widehat{\beta}_s} \; \bA^{-1}
  \frac{\partial \bA^{}}{\partial \widehat{\beta}_s}
   -
  \frac{\partial^2 \bA^{}}{\partial \widehat{\beta}_s^2} \;
  \right]
  \bA^{-1}
\end{eqnarray}
to allow for the computation of~$(a^{ss})''$.
To compute~(\ref{eq:deriv2.matrix.1param}) requires
\begin{eqnarray}
  \nonumber
  \frac{\partial^2 w_i}{\partial \widehat{\beta}_s^2}
  &~=~&
\left[
  \frac{\partial^2 w_i}{\partial \theta^2} \; 
  \frac{\partial \theta}{\partial \eta}
  +
  \frac{\partial w_i}{\partial \theta} \; 
  \frac{\partial^2 \theta}{\partial \eta^2}
  \frac{\partial \eta}{\partial \theta}
\right]
  \frac{\partial \theta}{\partial \eta}
\;
  x_{is}^2 \\
  &~=~&
\left[
  \frac{\partial^2 \, \EIMsub{i}}{\partial \theta^2}
  \left(
  \frac{\partial \theta}{\partial \eta}
  \right)^4
  + 4 \;
  \frac{\partial \, \EIMsub{i}}{\partial \theta} \;
  \frac{\partial^2 \theta}{\partial \eta^2}
  \left(
  \frac{\partial \theta}{\partial \eta}
  \right)^2
  +
  \mbox{} \right. \nonumber \\ && \left.
  2\; \EIMsub{i} \;
  \frac{\partial^3 \theta}{\partial \eta^3}
  \frac{\partial \theta}{\partial \eta}
  +
  \frac{\partial w_i}{\partial \theta} \;
  \frac{\partial^2 \theta}{\partial \eta^2}
\right]
  x_{is}^2 .
\label{eq:hdeff:d2w.dbetas2}
\end{eqnarray}
This in turn requires the
third derivatives of the link function
in its inverse form.
The ordinary form~$\partial^3 \eta / \partial \theta^3$ is
straightforward while the inverse form can be reexpressed as
\begin{eqnarray}
\frac{ \partial^3 \theta}{\partial \eta^3}
&~=~&
\left(
\frac{ \partial \theta}{\partial \eta}
\right)^{\! 4}
\left[
3 \;
\frac{ \partial \theta}{\partial \eta}
\left(
\frac{ \partial^2 \eta}{\partial \theta^2}
\right)^{\! 2}
-
\frac{ \partial^3 \eta}{\partial \theta^3}
\right],
\label{eq:Cwrte:linkfun.inverse.deriv3}
\end{eqnarray}
e.g.,
for logistic regression
$\partial^3 \eta / \partial \mu^3= 2\{1-3\mu(1-\mu)\}/[\mu(1-\mu)]^2$
and~(\ref{eq:Cwrte:linkfun.inverse.deriv3})
is~$\mu(1-\mu) \{ 1 - 6\mu(1-\mu) \}$.

\end{enumerate}

\section{Inference}
\label{sec:Cgelimo:hdeff.inference}

\subsection{Results for a 1-Parameter Regular Model}
\label{sec:Cgelimo:hdeff.lrt.1parameter}

The results for~$M=1$ models can be investigated further.
Consider the special case of~$\beta_s$ being the sole
parameter~$\theta$ of a regular distribution.
For simplicity use the observed information here.
We have the following result concerning two tipping points.

\noindent \textbf{Theorem~1.}
\textit{For~$H_0: \theta = \theta_0$
versus~$H_1: \theta \neq \theta_0$,
and
where the observed information
is evaluated at~$\widehat{\theta}$,
if the HDE is present then}
\begin{itemize}

\item[\textit{(a)}]  ~~~
\textit{the ratio of the Wald and LRT statistics satisfies}
\begin{eqnarray}
  \label{eq:lrt.1par.hde}
 \frac{ \calW  (\widehat{\theta},\ \theta_0) }{
        \calW_L(\widehat{\theta},\ \theta_0) }
&~<~&
   \frac35 + O_p(n^{-1}),
\end{eqnarray}

\item[\textit{(b)}] ~~~
\textit{and the ratio of the Wald and Rao score test statistics satisfies}
\begin{eqnarray}
  \label{eq:lrt.1par.hde.wald.rao}
 \frac{ \calW  (\widehat{\theta},\ \theta_0) }{
        \calW_S(\widehat{\theta},\ \theta_0) }
&~<~&
   \frac14 + O_p(n^{-1}).
\end{eqnarray}

\end{itemize}

\noindent \textbf{Proof}.\\
(a)
\qquad
As~$a^{11} = \left[ -\ell''(\widehat{\theta}) \right]^{-1}$
the HDE is present iff~(\ref{eq:hdeff:aberrant.condit}):
\begin{eqnarray}
  \label{eq:lrt.1par.hde.cond}
1
  &~<~&
\frac{\widehat{\theta} - \theta_0}{2} \cdot
\frac{ \displaystyle{\frac{d}{d\theta}}
       \left[ -\ell''(\widehat{\theta}) \right]^{-1}}{
       \left[ -\ell''(\widehat{\theta}) \right]^{-1}}
~=~
\frac{\widehat{\theta} - \theta_0}{2} \;
\frac{ \ell'''(\widehat{\theta})}{
       \left[ -\ell''(\widehat{\theta}) \right]}.
\end{eqnarray}
A Taylor series expansion of~$\ell(\theta_0)$
about~$\widehat{\theta}$ gives
\begin{eqnarray}
\calW_L(\widehat{\theta},\ \theta_0)
&~=~&
-\ell''(\widehat{\theta})
\;
\left( \widehat{\theta} - \theta_0 \right)^{2} +
\frac13 \, \ell'''(\widehat{\theta}) \;
\left( \widehat{\theta} - \theta_0 \right)^{3}
+ O_p(n^{-1})
\nonumber
\\
&~=~&
\left[ -\ell''(\widehat{\theta}) \right] \;
\left( \widehat{\theta} - \theta_0 \right)^{2}
\left\{
1 +
\frac23 \cdot
\frac12 \,
\frac{\ell'''(\widehat{\theta}) \;
       \left( \widehat{\theta} - \theta_0 \right)}{
      \left[ -\ell''(\widehat{\theta}) \right]}
\right\}
+ O_p(n^{-1})
\nonumber
\\
&~>~&
\left[ -\ell''(\widehat{\theta}) \right] \;
\left( \widehat{\theta} - \theta_0 \right)^{2}
\left[
1 +
\frac23 \cdot 1
\right]
+ O_p(n^{-1})
\mathrm{~~~by~}(\ref{eq:lrt.1par.hde.cond})
\nonumber
\\
&~=~&
\frac53 \;
\calW(\widehat{\theta},\ \theta_0)
+ O_p(n^{-1}).
\nonumber
\end{eqnarray}
The inequality~(\ref{eq:lrt.1par.hde})
follows from the property~$\calW_L = O_p(1)$.

\medskip

\noindent
(b) \qquad
Expand $\ell'(\theta_0)$ about the MLE:
\begin{eqnarray*}
  \ell'(\theta_0)
  &=&
  \ell'(\widehat{\theta}) +
  \ell''(\widehat{\theta}) (\theta_0 - \widehat{\theta}) +
  \frac12
  \ell'''(\widehat{\theta}) (\theta_0 - \widehat{\theta})^2 +
  \frac16
  \ell''''(\widehat{\theta}) (\theta_0 - \widehat{\theta})^3 +
  O_p(n^{-1})\\
  &=&
  j(\widehat{\theta}) (\widehat{\theta} - \theta_0) -
  \frac12
  j'(\widehat{\theta}) (\widehat{\theta} - \theta_0)^2 +
  \frac16
  j''(\widehat{\theta}) (\widehat{\theta} - \theta_0)^3 +
  O_p(n^{-1})\\
  &=&
  j(\widehat{\theta}) (\widehat{\theta} - \theta_0)
  \left[ 1 + Q + O_p(n^{-1}) \right]
\end{eqnarray*}
where
$Q := \frac12 (\theta_0 - \widehat{\theta})
\widehat{j'} / \widehat{j} > 1$
if and only if the HDE occurs.
Divide both sides by~$\sqrt{\widehat{j}}$ so that
\[
\widetilde{\calW}_s ~=~
\widetilde{\calW}
\left[ 1 + Q + O_p(n^{-1}) \right],
\]
i.e.,
\begin{eqnarray*}
  \frac{\widetilde{\calW}_s}{\widetilde{\calW}  }
  ~>~ 2 + O_p(n^{-1})
\end{eqnarray*}
when the HDE occurs.
Take the reciprocal
\begin{eqnarray*}
  \frac{\widetilde{\calW}  }{\widetilde{\calW}_s}
  &~<~&
  \frac12 + O_p(n^{-1})
\end{eqnarray*}
and square both sides to
obtain~(\ref{eq:lrt.1par.hde.wald.rao}).
\hfill{$\Box$}

Equation~(\ref{eq:lrt.1par.hde}) can be interpreted by saying
that if the Wald statistic becomes too small relative to the LRT
statistic (which is likely to be more accurate) then the HDE will
become present.  Likewise we can
interpret~(\ref{eq:lrt.1par.hde.cond}) by saying that
as~$|\widehat{\theta} - \theta_0| \rightarrow \infty$, if the
negative second derivative of~$\ell(\widehat{\theta})$ does not
grow fast enough compared to the the third derivative
of~$\ell(\widehat{\theta})$ then the HDE will become present.
The accuracy of the $3/5$ bound depends upon the fourth and
higher derivatives of~$\ell$.

Equations~(\ref{eq:lrt.1par.hde})
and~(\ref{eq:lrt.1par.hde.wald.rao}) suggest that
$\calW_L \approx \frac{5}{12} \calW_S$ when the HDE first starts
to occur.
In fact empirical findings indicate that
$\calW_L  / \calW_S \ll {5}/{12}$ in the presence of a strong HDE.

Extending the result to the two-parameter case is wieldy.
However, applying this result to the 2-parameter
\cite{hauc:donn:1977} data,
the 3/5 threshold lies between~$R=93$ and~94 successes,
whereas the HDE becomes present for~$\widehat{\beta}_2$
at $R \geq 92$
(Figure~\ref{fig:Cgelimo:hdonner.sims}(a)--(b)).
This suggests that the method can work well when the number
of parameters is low or are orthogonal.
As another numerical example,
when fitting a Poisson regression to the data described
for~(\ref{eq:Cgelimo:poissonff.dW2.dbeta2}) one obtains
a perfect match because the two cases of HDE present
corresponds to a ratio~$<3/5$ and the other cases
have a ratio~$>3/5$.

Unfortunately the LRT and Wald statistics are not independent;
if they were then their ratio would have a~$F_{1,1}$ distribution
whose mean is infinite.
The lower tail probability at the~$3/5$ quantile of this
distribution is~$0.42$, indicating that their positive
correlation creates a bias away from the null.

As~$\calW_L - \calW = O_p(1/\sqrt{n})$,
it follows that~$\calW / \calW_L= 1 + O_p(1/\sqrt{n})$.
Indeed, $\calW / \calW_L$ has
approximate asymptotic expectation~$3 - \Cov(\calW_L,\ \calW)$,
where
\begin{eqnarray*}
\Cov(\calW_L,\ \calW)
&~\approx~&
\Cov\!
\left(\calW_L,\ \calW_L - \frac13 \, \ell'''(\theta_0)
\left( \widehat{\theta} - \theta_0 \right)^{3}
            + O(n^{-2}) \right)
\\
&~\approx~&
\Var\,\calW_L - \frac13 \,
\Cov\!
\left( 2 \left[ \ell(\widehat{\theta}) - \ell(\theta_0) \right], ~
\ell'''(\theta_0)
\left( \widehat{\theta} - \theta_0 \right)^{3} \right)
\\
&~\approx~&
2 - \frac23 \,
  \ell'''(\theta_0) \;
\Cov\!
\left(
  \ell'(\theta_0) \left( \widehat{\theta} - \theta_0 \right), ~
  \left( \widehat{\theta} - \theta_0 \right)^{3} \right)
\\
&~=~&
2 - \frac23 \,
  \ell'''(\theta_0) \;
  \ell'(\theta_0) \;
  E \!
  \left[
  \left( \widehat{\theta} - \theta_0 \right)^{\! 4}
  \right]
  \\
&~\approx~&
2 \left[ 1 - \frac13 \,
  \ell'''(\theta_*) \;
  \ell'(\theta_0) \;
  3 \, \Var\!\left( \widehat{\theta} \right)^{2}
 \right],
\end{eqnarray*}
as~$\widehat{\theta} ~\approxsim~
N\left(\theta_0,\ [-\ell''(\theta_0)]^{-1} \right)$ under~$H_0$,
and~$\mu_4 = 3\sigma^4$ for a normal distribution.
Hence
\begin{eqnarray}
\Cov(\calW_L,\ \calW)
&~\approx~&
2 \left[
1 -  \ell'(\theta_0)  \cdot
\{ \ell''(\theta_0)  \}^{-2} \cdot
\ell'''(\theta_0)
\right].
\label{eq:cov.wald.stat.lrt.stat}
\end{eqnarray}
Thus
\begin{eqnarray}
E (\calW / \calW_L)  ~\approx~
E (\calW_L / \calW)
&~\approx~&
1  + 2\,
\ell'(\theta_0)  \cdot
\{ \ell''(\theta_0)  \}^{-2} \cdot
\ell'''(\theta_0),  ~~~~
\label{eq:E.wald.stat.lrt.stat}
\end{eqnarray}
and similarly
\begin{eqnarray*}
\Corr(\calW_L,\ \calW)
&~\approx~&
1 - \ell'(\theta_0)  \cdot
\{ \ell''(\theta_0)  \}^{-2} \cdot
\ell'''(\theta_0),
\label{eq:Cor.wald.stat.lrt.stat}
\\
\Var(\calW / \calW_L)  ~\approx~
\Var(\calW_L / \calW)
&~\approx~&
4 \,
\ell'(\theta_0)  \cdot
\{ \ell''(\theta_0)  \}^{-2} \cdot
\ell'''(\theta_0).
\end{eqnarray*}
An approximate upper bound from
Chebyshev's inequality shows that
\[
\Pr(|\calW/\calW_L - 1| \geq \frac25)  ~\leq~
\frac{\Var(\calW/\calW_L)}{(2/5)^2} ~=~
25 \,
\ell'(\theta_0)  \cdot
\{ \ell''(\theta_0)  \}^{-2} \cdot
\ell'''(\theta_0),
\]
as the probability of the HDE occuring by chance, given $H_0$,
however the bound is not very sharp.

A closing remark is that the asymptotic normality of the MLE can be
augmented with an additional regularity condition to prevent the HDE,
by restricting the parameter space.
Under~$H_0$, the extra condition is
\begin{eqnarray}
  \label{eq:lrt.1par.hde.cond.ell3}
\bTheta_* ~=~
\left\{
  \theta: ~
\frac{\theta - \theta_0}{-2}
  \cdot
\frac{\ell'''(\theta)}{\ell''(\theta)}
  ~<~ 1
\right\},
\end{eqnarray}
so that~$\sqrt{n}
\left( \widehat{\theta} - \theta_0 \right) ~\convergeD~
N(0,\ \EIMinvone(\theta_0))$
in~$\bTheta_*$ \citep[see, e.g.,][pp.294--5]{cox:hink:1974}.
Here, \EIMone{} is the expected information of one observation,
and \convergeD{} denotes convergence in distribution.

\subsection{Disproportional Sampling}
\label{sec:Cgelimo:hdeff.dispro}

In the case of a $2\times 2$ table it is now shown that, for a fixed
size effect~$\widehat{\beta}_2$, disproportional sampling can be used
to circumvent the HDE.

Table~\ref{tab:Cgelimo:hauc:donn:1977.dataset}(b)
is a modification of the HD data to allow for
disproportional sampling.
Here, $R \rightarrow N$ so that the~(2,\ 1) cell becomes low,
hence individuals with~$x_2=1$ need to be sampled with greater
intensity. This is achieved by having~$N_1^*=c^* N_1$
where the parameter~$c^* \geq 1$.
With~$c^* = 1$ being the usual scenario,
increasing~$N$ had no affect on the HDE
as the sign of~(\ref{eq:Cgelimo:waldstat.deriv1}) is unchanged.
The quantity~$f_0 = N_0 / N_1^*$ is then used to measure the relative
sampling intensity.  The sample
proportions~$\widehat{\pi}_0 = R_0 / N_0$
and~$\widehat{\pi}_1 = R / N_1$ remain unchanged.

With a logit link the new sampling scheme affects the intercept only.
The inverse crossproduct matrix is
\begin{eqnarray*}
\bA^{-1} ~=~
\frac{1}{N_1^*}
\left(
\begin{array}{cc}
  f_0^{-1} \, \widehat{\pi}_0^{-1} (1-\widehat{\pi}_0)^{-1} &
 -f_0^{-1} \, \widehat{\pi}_0^{-1} (1-\widehat{\pi}_0)^{-1} \\
 -f_0^{-1} \, \widehat{\pi}_0^{-1} (1-\widehat{\pi}_0)^{-1} &
~~f_0^{-1} \,
  \widehat{\pi}_0^{-1} (1-\widehat{\pi}_0)^{-1} +
  \widehat{\pi}_1^{-1} (1-\widehat{\pi}_1)^{-1}~
\end{array}
\right),
\end{eqnarray*}
and then by~(\ref{eq:deriv.matrix}),
$(a^{22})' =
({2 \, \widehat{\pi}_1 - 1}) / [{N_1^* \,
  \widehat{\pi}_1 \, (1-\widehat{\pi}_1)}]$.
The HDE will be present for~$\widehat{\beta}_2$ if
\begin{eqnarray}
1
&~<~&
\widehat{\beta}_2 \cdot
\left( \widehat{\pi}_1 - \frac12 \right) \;
\frac{
f_0 \,  \widehat{\pi}_0 (1-\widehat{\pi}_0)}{
f_0 \,  \widehat{\pi}_0 (1-\widehat{\pi}_0) +
        \widehat{\pi}_1 (1-\widehat{\pi}_1)}.
\label{eq:Cgelimo:hd77.HDE.dispro}
\end{eqnarray}
This
shows that~$\widehat{\beta}_2 > 2$ is needed before
  the HDE is possible, provided that~$\widehat{\pi}_1 \approx 1$
  and~$\widehat{\pi}_0$ is away from the boundaries.
  This corresponds to an odds ratio of about~7.4 or higher.
If $f_0 \approx 0$ then the HDE is unlikely,
in particular, the quantity
\[
\frac{f_0 \,  \widehat{\pi}_0 (1-\widehat{\pi}_0)}{
              \widehat{\pi}_1 (1-\widehat{\pi}_1)}
~~~(= \gamma, ~ \mathrm{say}),
\]
measures the sampling effect on the HDE:
small/large values of~$\gamma$ implies HDE is unlikely/likely
respectively.
This make intuitive sense as the (2, 1) cell increases
as a function of the total sample size
by choosing~$c^* \rightarrow \infty$ so that~$f_0 \rightarrow 0^+$.

Appendix~A6 extends the results of this section to a more
general setting.
In particular, when~$\pi_0 = \frac12$
in Table~\ref{tab:Cgelimo:hauc:donn:1977.dataset}(a)
so that~$\beta_1$
in~(\ref{eq:Cgelimo:hauc:donn:1977.model})
need not be estimated,
then it is shown that
approximately~$|\widehat{\beta}_2| > 2.40$ is needed
in order for the HDE to occur.
This corresponds to an odds ratio of about~$11.0$ or more,
or about~$0.091$ or less.

\section{Refinements}
\label{sec:Cgelimo:hdeff.refinements}

\begin{figure}[tt]
\begin{center}
\center
\resizebox{0.55\textwidth}{!}
{\includegraphics{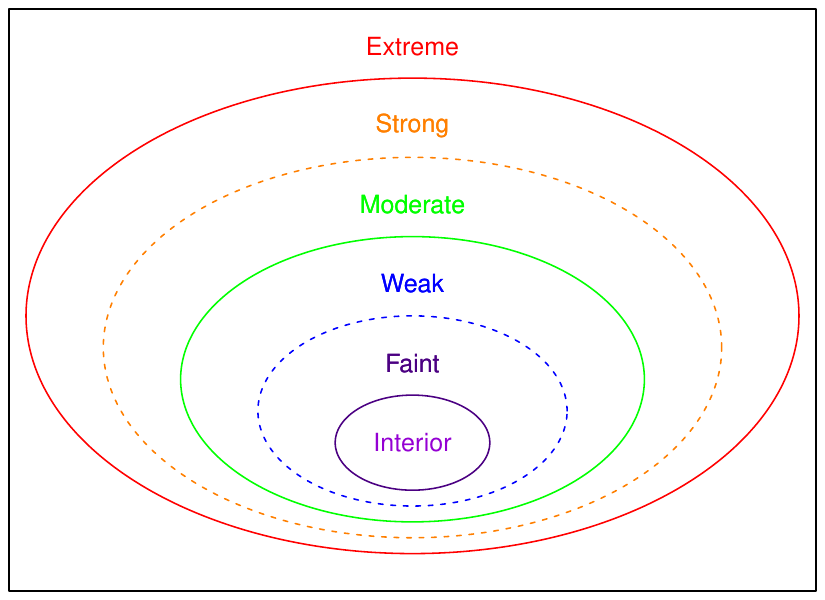}}
\caption{
Schematic diagram of the parameter space~\bTheta,
as characterized by HDE severity measures.
}
\label{fig:Cgelimo:hdeff.paramspace}
\end{center}
\end{figure}

\subsection{Finite-Differences}

Unfortunately implementing HDE detection for any particular model based
on~(\ref{eq:hdeff:dW.dbetas}) is labor-intensive,
e.g., Appendix~A1,
which has two consequences.
Firstly,
  this work suggests that the EIM is to be preferred over the
  observed information matrix
  (cf.\ \cite{efro:hink:1978} who prefer the latter),
  because terms often vanish upon expectation and
  therefore lead to simplification,
  e.g., (\ref{eq:uninormal}).
Secondly,
  numerical computation of the first two derivatives
  of~$\widetilde{\calW}$ circumvents this problem
  and provides a general method that
  empirical testing has shown to work well.
In particular, simplify~(\ref{eq:hdeff:dW.dbetas}) to
\begin{eqnarray}
  \frac{\partial \bW_i}{\partial \widehat{\beta}_{(r)s}^{*}}
  ~=~
  \sum_{j=1}^M \;
  \frac{\partial \bW_i}{\partial \eta_j}\;
  (\bH_s)_{jr} \;
  x_{isj}
  \label{eq:hdeff:dW.dbetas.simple}
\end{eqnarray}
where ${\partial \bW_i} / {\partial \eta_i}$
may be approximated by, e.g., central finite differences.
Similarly,
\begin{eqnarray}
  \frac{\partial^2 \bW_i}{\partial \widehat{\beta}_{(r)s}^{*2}}
  ~=~
  \label{eq:hdeff:d2W.dbetas2}
  \sum_{t=1}^M \;
  \sum_{j=1}^M \;
  \frac{\partial^2 \bW_i}{\partial \eta_t \, \partial \eta_j }\;
  (\bH_s)_{tr} \;
  (\bH_s)_{jr} \;
  x_{ist} \,
  x_{isj}
\end{eqnarray}
with separate formulas to handle the diagonal and
off-diagonal elements.
A step value of~$h_j \approx 0.005$ has been found to be reasonable
for most models, being on the~$\eta_j$-scale.

\subsection{Severity Measures and Parameter Space Partitioning}
\label{sec:Cgelimo:hdeff.severity}

So far, the above establishes whether a particular model suffers from
the HDE, however, a small negative derivative could indicate a mild or
extreme effect. It could also be argued that once the Wald statistic
starts to decrease but without having a negative derivative, then the
HDE has already started to occur.  Fortunately, it is possible to
gauge how severe the HDE is based the first two derivatives of
$\widetilde{\calW}_s$.
As $\widetilde{\calW}_s$ is asympotically
locally quadratic about the origin,
$\widetilde{\calW}_s(\widehat{\beta})$ is assumed piecewise
convex--concave--convex for~$\widehat{\beta}>0$, and
ditto for ~$\widehat{\beta}<0$.
Table~\ref{tab:hdework:severity} is a summary.

\begin{figure}[pp]
\begin{center}
\includegraphics[width=0.65\textwidth]{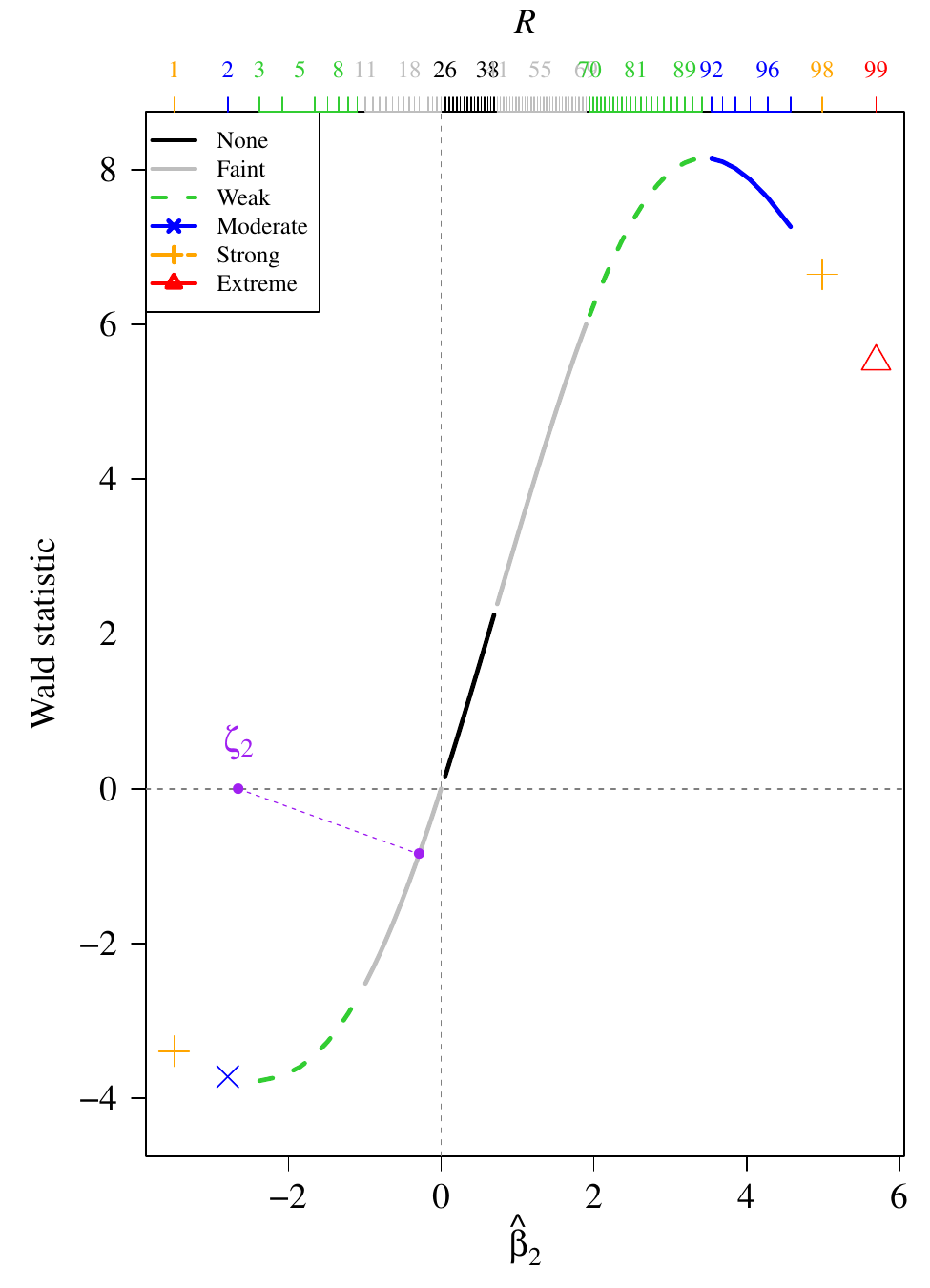}
\end{center}
\caption{
Color-coded HDE severity for the Hauck-Donner data
(Table~\ref{tab:Cgelimo:hauc:donn:1977.dataset}(a)
with~$R_0=25$),
cf.~Figure~\ref{fig:Cgelimo:hdonner.sims}(a).
The first derivative of the point~$\zeta_2$,
which is the intersection of the normal line
at~$(\widehat{\beta}_{2}, \widetilde{\calW}(\widehat{\beta}_{2}))$
with the $x$-axis,
is used to categorize sections of the curve.
\label{fig:Cgelimo:hdonner.severity}
}
\end{figure}

Using the simple notation~$\widehat{\beta}$ to denote the $x$-axis,
let~$\zeta(\widehat{\beta})$ be
the intersection of the normal line at
$(\widehat{\beta}_{}, \widetilde{\calW})$ with the $x$-axis
(purple dashed line of Figure~\ref{fig:Cgelimo:hdonner.severity}).
The movement of~$\zeta$ as the coefficient changes allows
further properties of the curve
to be determined
additional to the location of inflection points.
(The use of~$\zeta$ is a more convenient alternative to using the
tangent.)
Denoting the cutpoints as
$\widehat{\beta}_{nf}$,
$\widehat{\beta}_{fw}$,
$\widehat{\beta}_{wm}$,
$\widehat{\beta}_{ms}$,
$\widehat{\beta}_{se}$
for
\textit{n}one,
\textit{f}aint,
\textit{w}eak (mild),
\textit{m}oderate,
\textit{s}evere (strong),
\textit{e}xtreme,
they are defined by
$\widetilde{\calW}''(\widehat{\beta}_{nf}) = 0$,
$\zeta'(\widehat{\beta}_{fw}) = 0$,
$\widetilde{\calW}'(\widehat{\beta}_{wm}) = 0$,
$\zeta'(\widehat{\beta}_{ms}) = 0$,
$\widetilde{\calW}''(\widehat{\beta}_{se}) = 0$,
so that
$\widehat{\beta}_{nf}$ and
$\widehat{\beta}_{se}$ are inflection points.
For positive estimates
\begin{eqnarray*}
  \label{eq:hdeff.severity.ordinal}
0 \leq
\widehat{\beta}_{nf} \leq
\widehat{\beta}_{fw} \leq
\widehat{\beta}_{wm} \leq
\widehat{\beta}_{ms} \leq
\widehat{\beta}_{se} < \infty,
\end{eqnarray*}
and
\begin{eqnarray}
  \label{eq:hdeff.severity.dzeta}
\zeta(\widehat{\beta}_{})
  &=&
\widehat{\beta}_{} +
        \widetilde{\calW} (\widehat{\beta}_{})  \cdot
        \widetilde{\calW}'(\widehat{\beta}_{}),
\\
\zeta'(\widehat{\beta}_{})
  &=&
1 +
\left\{ \widetilde{\calW}'(\widehat{\beta}_{}) \right\}^2 +
        \widetilde{\calW}(\widehat{\beta}_{})   \cdot
        \widetilde{\calW}''(\widehat{\beta}_{}).
 \nonumber
\end{eqnarray}

For the data set
of Table~\ref{tab:Cgelimo:hauc:donn:1977.dataset}(a)
the scheme classifies
no HDE for~$R=26,\ldots,40$,
faint HDE for~$R=11,\ldots,25,41,\ldots,69$,
mild HDE for~$R=3,\ldots,10,70,\ldots,91$,
moderate HDE for~$R=2,92,\ldots,97$,
severe HDE for~$R=1,98$, and
extreme HDE for~$R=99$
(Fig.~\ref{fig:Cgelimo:hdonner.severity}).

\begin{table}[tt]
\caption{
HDE severity measures:
the~0s define~5 cutpoints
that are used to define 6~categories of HDE.
The sign of various quantities are given where possible, and
the key property of $\widetilde{\calW}_s$ is given.
Approximate boundary values of~$R$ are for the data set of
Table~\ref{tab:Cgelimo:hauc:donn:1977.dataset}(a)
with~$R_0=25$.
\label{tab:hdework:severity}
}
\centering
\begin{tabular}{llcccc}
  \hline
Severity  ~~~~~~ &
Key property ~~ &
~~ $\widetilde{\calW}_s'$ ~~ &
~~ $\sgn(\widehat{\beta}_s) \cdot \widetilde{\calW}_s''$ ~~ &
~~ $\zeta'$ ~~ &
~~ $R$ ~~
  \\
  \hline
None
& convex
              & $+$ & $+$
              & $+$ &  $    $ \\
{[boundary]} & & $+$ & $0$ & $+$ &       $25.5,\ 40.5$ \\
Faint (Very mild) &
concave & $+$ & $-$ & $+$ &       $    $ \\
{[boundary]} & & $+$ & $-$ & $0$ &       $10.5,\ 69.5$ \\
Weak (Mild) &
concave & $+$ & $-$ & $-$ &       $    $ \\
{[boundary]} &  & $0$ & $-$ & $-$ &       $2.5,\ 91.5  $ \\
Moderate    &
concave & $-$ & $-$ & $-$ &       $    $ \\
{[boundary]} & & $-$ & $-$ & $0$ &       $1.5,\ 97.5$ \\
Strong (Severe) ~~~ &
concave & $-$ & $-$ & $+$ &       $    $ \\
{[boundary]} & & $-$ & $0$ & $+$ &       $0.5,\ 98.5$ \\
Extreme  (Very severe) ~~   &
convex & $-$ & $+$ & $+$ &       $    $ \\
   \hline
\end{tabular}
\end{table}

As a consequence, the HDE severity measures allow the parameter space
to be partitioned \textit{practically} into an interior where the
regularity conditions hold, which is encased by layers at the boundary
of increasing HDE severity.  Figure~\ref{fig:Cgelimo:hdeff.paramspace}
displays this schematically.  For a given data set and regression
model, not all the layers present and \bTheta{} may be discrete. The
descriptors should be interpreted relatively rather than
absolutely---their purpose is to provide an ordinal categorization of
the HDE severity.

\section{Examples}
\label{sec:Cgelimo:hdeff.eg}

\subsection{Birds and Logistic Regression}

\cite{mang:2015} applies a variable selection algorithm
based on the AIC to choose covariates in a logistic
regression applied to a bird data set
with 67 cases after missing values have been removed.
Six variables plus an intercept were chosen
out of a possible~13 variables
(Table~\ref{tab:birds.waldtable})
for the response \texttt{Status}.

For this model, 3 variables display the HDE: \texttt{Upland},
\texttt{Migr} and \texttt{Indiv} (Table~\ref{tab:birds.waldtable}),
and \texttt{Mass} is weakly affected.  According to the usual
Wald test, the former two have p-values of moderate strength:
2\% and 3\%, while \texttt{Indiv} is more strongly significant
(0.005).  However, LRT p-values for the 3 variables are 0.00135,
0.0076, $5\times10^{-11}$ respectively.  This suggests that
these variables are considerably more statistically significant
than would na\"{i}vely appear; the ratios of the p-values are
approximately 15, 4 and $1 \times 10^{7}$.  The latter is huge
and shows that a moderate HDE can be associated with an enormous
relative effect on the p-value (although qualitatively similar);
that variable is actually skewed and a log-transformation would
be recommended.

\begin{table}[tt]
\caption{
Logistic regression fitted to a birds data set:
the Wald table plus the first derivative of the
signed root Wald statistics,
as well as the first two derivatives of the
standard errors.
The HDE severity is included.
All values are rounded to 2 decimal places.
Here, $s=1,\ldots,7$ indexes the regression coefficients in order.
\label{tab:birds.waldtable}
}
\centering
\begin{tabular}{rrrrrcrr}
  \hline
Parameter & ~~MLE & ~~~SE   & $\widetilde{\calW}_s$
  & ~ $\widetilde{\calW}_s'$
  & ~HDE severity~
  & ~~ $(\sqrt{a^{ss}})'$
  & ~~ $(\sqrt{a^{ss}})''$
  \\
  \hline
\texttt{(Intercept)} &
$-$3.44 & 2.06 & $-$1.67 & 0.03 & Faint & $-$0.56 & 0.18 \\
\texttt{Upland} &
$-$4.73 & 2.04 & $-$2.32 & $-$0.05 & Strong & $-$0.47 & 0.13 \\
\texttt{Migr} &
$-$2.02 & 0.93 & $-$2.17 & $-$0.05 & Moderate & $-$0.48 & 0.33 \\
\texttt{Mass} &
0.00 & 0.00 & 2.73 & 186.40 & Weak & 0.32 & 559.30 \\
\texttt{Indiv} &
0.01 & 0.00 & 3.48 & $-$46.19 & Moderate & 0.34 & 35.03 \\
\texttt{Insect} & 0.27 & 0.14 & 1.90 & 7.87 & Faint & $-$0.06 & 1.38 \\
\texttt{Wood} & 1.95 & 1.30 & 1.50 & 0.75 & Faint & 0.02 & 0.06 \\
   \hline
\end{tabular}
\end{table}

\subsection{Wine and the Partial Proportional Odds Model}

Table~1 of \cite{kosm:2014}
reports the Wald table of a partial
proportional odds model (PPOM) fitted to a wine tasting data set where the
response bitterness was measured on a 5-level ordinal scale.  The
two binary variables $x_2=$ `temperature' and
$x_3=$ `contact' were measured in this experiment, and
the parallelism assumption was applied to~$x_3$ only.
Three of the SEs are inflated.
The data was first analyzed in \cite{rand:1989}.

Table~\ref{tab:wine.waldtable} provides additional information to
the Wald table, viz.~the first two derivatives of the
signed root Wald statistics.
(The sign of some of the coefficients differ from his table
because the \textit{reversed} PPOM,
$\logit \, \Pr(Y \geq j+1) =\eta_j$, was used so that the direction
matches logistic regression.)
The HDE is evident in 3 of the regression coefficients:
$\widehat{\beta}_{(3)1}^*$,
$\widehat{\beta}_{(4)1}^*$,
$\widehat{\beta}_{(1)2}^*$,
where the latter two have the extreme form.
Furthermore,
$\widehat{\beta}_{(1)1}^*$ and
$\widehat{\beta}_{(2)2}^*$ are weakly affected.
The causes of the HDE in the first two, which relate to the intercepts,
may be confounded with
the requirement of satisfying the ordering
$\widehat{\beta}_{(1)1}^* \geq
 \widehat{\beta}_{(2)1}^* \geq
 \widehat{\beta}_{(3)1}^* \geq
 \widehat{\beta}_{(4)1}^*$.
For~$x_2$ the SE is so inflated that one of its
Wald statistics' slopes is only slightly negative.
For this,
the LRT p-value for testing~$H_0: {\beta}_{(1)2}^* = 0$ is 0.006,
which differs totally from the Wald test p-value of 1.00!

\cite{kosm:2014} discusses adjustments specifically for cumulative
link models and proposes bias reduction methods in order to safeguard
against infinite parameter estimates.
In this example, the number of coefficients manifesting
the HDE ought to alert the practitioner of a nonstandard
situation.
Furthermore,
suspicion for something gone awry should have been raised by
monitoring convergence and observing
that this took~19 IRLS iterations---this is
considerably more than the usual 6--8 iterations---and that the
decrease in deviance during the last 10 iterations was only slight.

\begin{table}[tt]
\caption{
Partial proportional odds model fitted to the wine tasting data set:
the Wald table plus the first two derivatives of the
signed root Wald statistics
and the HDE severity.
All values are rounded to 2 decimal places.
\label{tab:wine.waldtable}
}
\centering
\begin{tabular}{crrrrrc}
  \hline
Parameter & MLE & SE   & $\widetilde{\calW}_s$
  & $\widetilde{\calW}_s'$ & $\widetilde{\calW}_s''$
& ~~HDE severity~~ \\
  \hline
$\beta_{(1)1}^*$ & 1.27 & 0.51 & 2.48 & 1.25 & $-$1.32 & Weak  \\
$\beta_{(2)1}^*$ & $-$1.10 & 0.44 & $-$2.51 & 1.92 & 0.86 & Faint \\
$\beta_{(3)1}^*$ & $-$3.77 & 0.80 & $-$4.71 & $-$0.44 & 0.79 & Moderate \\
$\beta_{(4)1}^*$ & $-$23.23 & 6904.18 & $-$0.00 & $-$0.00 & $-$0.00 &
Extreme \\
$\beta_{(1)2}^*$ & 21.53 & 11433.17 & 0.00 & $-$0.00 & 0.00 & Extreme \\
$\beta_{(2)2}^*$ & 2.15 & 0.59 & 3.65 & 1.02 & $-$0.94 & Weak \\
$\beta_{(3)2}^*$ & 2.87 & 0.81 & 3.54 & 1.22 & $-$0.15 & Faint \\
$\beta_{(4)2}^*$ & 20.89 & 6904.18 & 0.00 & 0.00 & $-$0.00 & Faint \\
$\beta_{(1)3}^*$ & 1.47 & 0.47 & 3.13 & 1.87 & $-$0.15 & Faint \\
   \hline
\end{tabular}
\end{table}

\section{Computational Details  and Recommendations}
\label{sec:Cgelimo:hdeff.summary}

One can obtain an HDE-free Wald test
by evaluating the denominator of the Wald statistic
at~$\theta_{s0}$ rather than~$\widehat{\theta}_{s0}$.
For simplicity of notation,
this section enumerates the regression coefficients to be estimated
of a general VGLM as $\beta_k$ for $k=1,\ldots,p_{\stVLM}$.
The tests are
$H_{0k}: \beta_k = \beta_{0k}$ versus
$H_{1k}: \beta_k \neq \beta_{0k}$
where usually~$\beta_{0k}=0$.
By an ``HDE-free Wald test'', it is meant that~$\widehat{\beta}_k$
is replaced by its hypothesized value~${\beta}_{0k}$ when
computing the test statistic's SE so that its derivative
with respect to~$\widehat{\beta}_k$ vanishes.
Then two options open up: whether or not to perform IRLS iterations
for the other coefficients.
If so, then this is equivalent to fitting the LRT model under~$H_{0k}$,
so it is not surprising that the time cost is similar.
Notationally, let~$(\beta_{0k},\ \widehat{\bbeta}_{-k})$
correspond to no (further) iteration
and~$(\beta_{0k},\ \widehat{\widehat{\bbeta}}_{-k})$ be
with iteration.

Computationally, the use of $\beta_{0k}$ instead
of~$\widehat{\beta}_{k}$ is implemented by deleting the~$k$th column
of~$\bX_{\stVLM}$ and adding $x_{ik} \beta_{0k}$ to the matrix of
offsets.  If iterating for the other coefficients, initial values can
be obtained from the original model and then convergence is usually rapid
(often only 1 or 2 IRLS iterations are needed).

To obtain the SE for~$\beta_k$ for a HDE-free Wald test,
the main steps are as follows.
\begin{enumerate}

\item[(i)]
  Optionally iterate
  (i.e., compute~$\widehat{\widehat{\bbeta}}_{-k}$),
  else use the MLEs~$\widehat{\bbeta}_{-k}$ for the following.

\item[(ii)]
Compute~$\boldeta_i$ using~$(\beta_{0k},\ \widehat{\bbeta}_{-k})$
or~$(\beta_{0k},\ \widehat{\widehat{\bbeta}}_{-k})$.

\item[(iii)]
Update the generic fitted values~$\bmu_i$
and the working weights~$\bW_i$ from~$\boldeta_i$.

\item[(iv)]
Compute the Cholesky decompositions~$\bU_i$ of the~$\bW_i$.

\item[(v)]
Compute~$\Diag(\bU_1,\ldots,\bU_n) \, \bX_{\stVLM}$
and its QR-decomposition.

\item[(vi)]
Compute~$\bR^{-1}$ and then~$(\bR^{-1} \bR^{-T})_{kk}^{1/2}$ to
obtain~SE${}_k$.

\end{enumerate}
These steps are performed for each~$\beta_k$.

A small numerical study involving timing various VGLMs fitted with the
author's software gave the following results.
\begin{itemize}

\item
The cost of conducting
an HDE test on all the regression coefficients is typically
about $\frac13$ the cost of obtaining the HDE-free
iterated Wald statistics.

\item
Without iterations, HDE-free
Wald tests can be about 25\% less costly time-wise compared to LRTs.

\item
With iteration,
HDE-free Wald tests can be about 30\% more costly time-wise
than LRTs.

\item
The cost of a score test is similar to an iterated Wald test.

\end{itemize}
These results suggest that the cost of a LRT and a HDE-free Wald test
is roughly comparable.

For the practitioner,
the above suggests that a reasonable strategy is to
firstly apply an HDE test to all the regression coefficients.
If any are affirmative,
then the next step is applied to those coefficients
and depends on whether
SEs are sought, e.g., as a rough measure of statistical uncertainty.
Thus, secondly,
when SEs are required, HDE-free Wald tests should be conducted
for those coefficients,
and should only $p$-values be necessary, then LRTs should be computed
instead.
As for deciding whether iterations are required,
if the computational expense is of concern, then
the non-iterated variant is suggested as it can be approximately
half the cost of the iterated variant.

It should be noted that conducting hypothesis tests may be difficult
for some models, such as a non-parallel cumulative link model because
of intrinsic order restrictions such as $\beta_{(1)k} < \cdots <
\beta_{(j)k} < \cdots < \beta_{(M)k}$.  Circumventing the difficulties
involved is an area for future research.
Another note is that the score test lacks the intuitive appeal of the
Wald test and may be inconsistent \citep{free:2007}, therefore it is
suggested that
the order of preference of the tests be, in decreasing order,
LRTs followed by HDE-free Wald tests followed by score tests.

\section{Discussion}

In an era of high-dimensional statistics and Big Data the $p$-value
remains a chief centerpiece of classical frequentist statistical
inference (\cite{deze:buhl:meie:mein:2015},
 \cite{fan:2014}, \cite{mein:meie:buhl:2009})
despite recent statements about their misuse \cite{wass:laza:2016}.
Indeed, \cite{sieg:2010} writes: ``It's science's dirtiest secret:
The `scientific method' of testing hypotheses by statistical
analysis stands on a flimsy foundation.''  Although most of
the weakness is interpretative this article highlights another
deficiency in the form of the HDE.

This work was motivated by practical Wald testing in a general
regression setting, and has developed methods for testing whether
an estimate $\widehat{\bbeta}$ gives rise to parameters so close
to the parameter space boundary that some Wald statistics are
aberrant.  The practical implication is clear: for SEs computed
at the MLE establishing whether the HDE is manifest in a fitted
model should be determined where possible, and if so then the LRT
or other large-sample tests be conducted instead.  Ideally HDE
testing should be carried out whenever a Wald table is presented,
and statistical software should be upgraded to make this
practical and automatic.  Software for variable selection based
on the usual Wald statistics need modification too.  With
business intelligence software companies being a major driver of
Big Data, it is plausible that litigation could occur if they
fail to respond adequately to implement new methodologies such as
here to address known flaws such as the HDE.  As a minimum,
disclaimer statements that the HDE is not detected could be
issued as interim measures.

This article is a first step towards shedding light on
the structure of the parameter space in a practical sense.  There
is further work to be done, such as refining the severity
measures of Section~\ref{sec:Cgelimo:hdeff.severity}, exploring
consequences under a Bayesian framework, and seeing if
adjustments are needed for multiplicity if we switch to another
test. Some new results can be found
in \cite{yee:2021.hde}.
This manuscript has appeared in published form
as \cite{yee:2022.jasa}.

A parting remark is that it has been disappointing to see that
most texts on statistical inference and practice do not even
mention the HDE, especially those making heavy use of Wald
tables.  Of the few that do, various authors have described the
HDE as `the major statistical problem of \calW{}' and `certainly
disturbing'.  It is hoped that this work will help make the HDE a
more recognized problem and provide a practical solution.

\bigskip

\noindent
\textbf{Acknowledgements} \qquad
I wish to thank the Centre for Applied Statistics and School of
Mathematics and Statistics at the University of Western Australia
for hospitality during a workshop given there in early 2017 and for
valuable feedback that helped lead to this work.  Helpful feedback
from George Seber and Elbert Chia is gratefully acknowledged.

\bigskip

\begin{center}
{\large\bf Appendices}
\end{center}

\noindent Notes:
\begin{description}

\item[\pkg{VGAM}]
  Version 1.0-4 introduced \texttt{hdeff()}    and appeared on CRAN in July 2017.

\item[\pkg{VGAM}]
  Version 1.1-1 introduced \texttt{hdeffsev()} and appeared on CRAN in Feb 2019.

\end{description}

\subsection*{A1. Two Working Weights Examples}

As two simple examples of~(\ref{eq:hdeff:dW.dbetas}),
the following drop the subscript~$i$ simplicity.
\begin{enumerate}

\item[(i)]
For the zero-inflated Poisson distribution parameterized by
\begin{eqnarray}
\Pr(Y=y;\phi,\lambda) ~=~
I_{[y=0]} \, \phi + (1- \phi) \, {e^{-\lambda} \lambda^y}/{y!}, ~~~~
y=0,1,\ldots, ~~~~
\label{eq:hdeff:dzipoisson.copy2}
\end{eqnarray}
with mixing probability~$\phi$
the two derivative matrices are
\begin{eqnarray*}
\frac{\partial \, \EIM{}}{\partial \phi}
&~=~&
\left(
\begin{array}{cc}
\displaystyle{\frac{-(1-e^{-\lambda})
         (1-2 \pi_0)}{
         (1-\phi)^2 \, \pi_0^2}} &
\displaystyle{\frac{e^{-\lambda} \, (1 - e^{-\lambda})}{
              \pi_0^2}} \\[1.2em]
\displaystyle{\frac{e^{-\lambda} \, (1 - e^{-\lambda})}{
              \pi_0^2}} &
\displaystyle{\frac{-1}{\lambda} -
\frac{e^{-\lambda}
\left[ (1-\phi)^2\, e^{-\lambda} - \phi^2 \right]}{\pi_0^2}
}
\end{array}
\right)
\label{eq:hdeff:deriv.phi.eim.zipoisson}
\end{eqnarray*}
and
\begin{eqnarray*}
\frac{\partial \, \EIM{}}{\partial \lambda}
&~=~&
\left(
\begin{array}{cc}
\displaystyle{\frac{e^{-\lambda}}{(1-\phi) \, \pi_0^2}} &
\displaystyle{\frac{\phi \, e^{-\lambda}}{\pi_0^2}} \\[1.2em]
\displaystyle{\frac{\phi \, e^{-\lambda}}{\pi_0^2}} &
\displaystyle{\frac{-(1-\phi)}{\lambda^2} +
\frac{ \phi^2 (1-\phi)\, e^{-\lambda}}{\pi_0^2}}
\end{array}
\right),
\label{eq:hdeff:deriv.lambda.eim.zipoisson}
\end{eqnarray*}
where~$\pi_0 = \phi+(1-\phi)\, e^{-\lambda}$ is the probability
of an observed~0.
When implementing these in software, symbolic differentiation may be
useful for higher order derivatives.

\item[(ii)]
For the cumulative link model
$g(\Pr(Y \leq j))= \eta_j$
taking levels~$\{1,\ldots,M+1\}$,
let the cumulative probabilities be~$\gamma_j$,
then its EIM is tridiagonal and yields
non-zero elements of its derivative matrix centered at
the $(s,s)$th element as
\begin{eqnarray*}
\frac{\partial \, \EIM{}}{\partial \gamma_s}
&~=~&
{N}
\left(
\begin{array}{ccc}
  \displaystyle{-\mu_{s  }^{-2}} &
  \displaystyle{ \mu_{s  }^{-2}} &
  0 \\
  \displaystyle{ \mu_{s  }^{-2}} &
~~\displaystyle{ \mu_{s+1}^{-2}} -
  \displaystyle{ \mu_{s  }^{-2}}~~ &
  \displaystyle{-\mu_{s+1}^{-2}} \\
  0 &
  \displaystyle{-\mu_{s+1}^{-2}} &
  \displaystyle{ \mu_{s+1}^{-2}}
\end{array}
\right), ~~~~ \mathrm{for}\ s=2,\ldots,M-1,
\end{eqnarray*}
with straightforward truncation of rows and columns for~$s=1$ and~$M$.
The second derivative matrices are
\begin{eqnarray*}
\frac{\partial^2 \, \EIM{}}{\partial \gamma_s^2}
&~=~&
{2\,N}
\left(
\begin{array}{ccc}
  \displaystyle{ \mu_{s  }^{-3}} &
  \displaystyle{-\mu_{s  }^{-3}} &
  0 \\
  \displaystyle{-\mu_{s  }^{-3}} &
~~\displaystyle{ \mu_{s+1}^{-3}} +
  \displaystyle{ \mu_{s  }^{-3}}~~ &
  \displaystyle{-\mu_{s+1}^{-3}} \\
  0 &
  \displaystyle{-\mu_{s+1}^{-3}} &
  \displaystyle{ \mu_{s+1}^{-3}}
\end{array}
\right), ~~~~ \mathrm{for}\ s=2,\ldots,M-1,
\end{eqnarray*}

\end{enumerate}
It is possible to use symbolic differentiation,
such as \RR{}'s \texttt{deriv3()},
however it is not very efficient
and produce numerically unstable estimates;
it fails to be a general purpose method.

\subsection*{A2. Sandwich Estimators}

Extensions to handle sandwich estimators are available,
e.g.,
for GLMs this is ($\bSigma$, say)
\begin{eqnarray}
  \label{eq:hdeff:sandwich.GLMs}
\left( \bX^T \widehat{\bW} \bX \right)^{-1}
\left[ \;
\bX^T \widetilde{\bW} \bX
\right]
\left( \bX^T \widehat{\bW} \bX \right)^{-1}
  &~=~&
\bA(\widehat{\bbeta})^{-1} \;
\bB(\widehat{\bbeta}) \;
\bA(\widehat{\bbeta})^{-1},   ~~~~~
\end{eqnarray}
where~$\widetilde{\bW}$ is diagonal with elements
$\left[ (y_i-\widehat{\mu}_i) \cdot
(\partial \mu_i / \partial \eta_i) \, \widehat{\phi}
/ V(\widehat{\mu}_i) \right]^2$
and variance
$V(\mu)$.
Then
\begin{eqnarray}
  \nonumber
  \frac{\partial \bSigma}{\partial \widehat{\beta}_s}
  &~=~&
  \frac{\partial \bA^{-1}}{\partial \widehat{\beta}_s} \;
   \bB^{} \;
   \bA^{-1}
  +
   \bA^{-1} \;
  \frac{\partial \bB^{}}{\partial \widehat{\beta}_s} \;
   \bA^{-1}
  +
   \bA^{-1} \;
   \bB^{} \;
  \frac{\partial \bA^{-1}}{\partial \widehat{\beta}_s}
  \\
  &~=~&
  \bA^{-1}
  \left[ \,
  \frac{\partial \bB^{}}{\partial \widehat{\beta}_s}
  -
   \frac{\partial \bA^{}}{\partial \widehat{\beta}_s} \;
   \bA^{-1} \;
   \bB^{}
  -
   \bB^{} \;
   \bA^{-1} \;
  \frac{\partial \bA^{}}{\partial \widehat{\beta}_s} \,
  \right]
   \bA^{-1}
  \label{eq:deriv.3matrices}
\end{eqnarray}
by~(\ref{eq:deriv.matrix})
so that~$\sigma_{ss}^{'}$ can be computed.

For logistic regression the diagonal elements of~$\widetilde{\bW}$
are~$(y_i - \widehat{\mu}_i)^2$
so that
\[
\frac{\partial \bB^{}}{\partial \widehat{\beta}_s} ~=~
-2 \sum_{i=1}^n \,
(y_i - \widehat{\mu}_i) \;
\widehat{\mu}_i \, (1 - \widehat{\mu}_i) \, x_{is} \,
\bix_i \, \bix_i^T.
\]
Applied to the Hauck \& Donner data, it is easily shown that
the ordinary SEs and sandwich estimators coincide (i.e.,
$\bSigma=\bA^{-1}$) because $\bA=\bB$.

\subsection*{A3. Multiple Tests}

Up till now simple null hypotheses of the form~$H_0: \beta_s = 0$
have been considered.
More generally, suppose we wish to test~$H_0: \bL \bbeta = \bic$
for some~$q \times p_{\stVLM}$ matrix~\bL{}
of rank~$q$ comprising known fixed constants,
and $q$-vector~\bic{} of known fixed constants.
With the usual regularity conditions holding the Wald statistic
\begin{eqnarray}
  \label{eq:hdeff:W2.stat.multiple.testing}
\calW ~=~
  \left( \bL \widehat{\bbeta} - \bic \right)^T \,
  \left(
  \bL{} \, \bA^{-1} \, \bL^T
  \right)^{-1}
  \left( \bL \widehat{\bbeta} - \bic \right)
\end{eqnarray}
is asymptotically~$\chi_q^2$ under~$H_0$.
To detect any HDE in~(\ref{eq:hdeff:W2.stat.multiple.testing})
let~$\widehat{\bdelta} = \bL \widehat{\bbeta} - \bic$,
and we conclude that the test suffers from HDE degradation if
${\partial \calW}/{\partial \widehat{\delta}_u} < 0$
for any component~$u \in \{1,\ldots,q\}$ of~$\widehat{\bdelta}$.
The handling of~(\ref{eq:hdeff:W2.stat.multiple.testing}) follows
from the main treatment of the paper but with the additional
computation
\begin{eqnarray*}
\lefteqn{
  \frac{\partial \calW}{\partial \widehat{\bdelta}_{}} ~=~
  2
  \left( \bL{} \, \bA^{-1} \, \bL^T \right)^{-1}
  \widehat{\bdelta} +
      \mbox{}} \\
&&
  \sum_{u=1}^{q}
  \left\{
  \widehat{\bdelta}^T
  \left( \bL{} \, \bA^{-1} \, \bL^T \right)^{-1}
  \bL \, \bA^{-1} \,
    \frac{\partial \bA}{\partial \widehat{\delta}_u} \,
  \bA^{-1} \, \bL^T
  \left( \bL{} \, \bA^{-1} \, \bL^T \right)^{-1}
  \widehat{\bdelta}
  \right\}
  \bie_u
\end{eqnarray*}
subject to
\[
\frac{\partial \bA^{-1}}{\partial \widehat{\beta}_s} ~=~
\sum_{u=1}^{q}
\frac{\partial \bA^{-1}}{\partial \widehat{\delta}_u}\,
(\bL)_{us}
\]
where the elements of~\bbeta{} are enumerated by~$\beta_s$.
This admits the solution
\begin{eqnarray}
  \sum_{u=1}^{q}
\left(
 \bie_u \otimes
 \frac{\partial \bA^{-1}}{\partial \widehat{\bdelta}_{u}}
\right) ~=~
\label{eq:hdeff:W2stat.multiple.testing.bdelta}
\sum_{s=1}^{p_{\mathrm{VLM}}}
\left[
\left( \bL \, \bL^T \right)^{-1} \bL \, \bie_s
\right]
\otimes
\frac{\partial \bA^{-1}}{\partial \widehat{\beta}_{s}}. ~~~
\end{eqnarray}

A simple example is $\bL=(1,-1)$ which yields
\[
\frac{\partial \bA^{-1}}{\partial \widehat{\delta}_1} ~=~
\frac12
\left(
  \frac{\partial \bA^{-1}}{\partial \widehat{\beta}_1} -
  \frac{\partial \bA^{-1}}{\partial \widehat{\beta}_2}
\right).
\]
Another simple example, relevant for a partial proportional
odds model, is
\[
\bL ~=~
\left(
\begin{array}{ccc}
\ 1~  &  ~0~ & ~-1~\\
\ 0~  &   1  & ~-1~
\end{array}
\right)
\]
which results in
\[
\left(
\begin{array}{c}
\displaystyle{
\frac{\partial \bA^{-1}}{\partial \widehat{\delta}_1}
  }\\
\displaystyle{
\frac{\partial \bA^{-1}}{\partial \widehat{\delta}_2}
}
\end{array}
\right) ~=~
\left(
\begin{array}{c}
\displaystyle{
\ \,\frac23}\,
\displaystyle{
\frac{\partial \bA^{-1}}{\partial \widehat{\beta}_1}
  }-
 \frac13 \,
\frac{\partial \bA^{-1}}{\partial \widehat{\beta}_2} -
\displaystyle{
 \frac13
  }\,
\frac{\partial \bA^{-1}}{\partial \widehat{\beta}_3}\\
\displaystyle{
 -\frac13
  }\,
\frac{\partial \bA^{-1}}{\partial \widehat{\beta}_1} +
\displaystyle{
 \frac23
  }\,
\frac{\partial \bA^{-1}}{\partial \widehat{\beta}_2} -
\displaystyle{
 \frac13
  }\,
  \displaystyle{
\frac{\partial \bA^{-1}}{\partial \widehat{\beta}_3}
  }
\end{array}
\right).
\]

\subsection*{A4. The Proportional Hazards Model}

The Cox model suffers potentially from the HDE
\citep[p.60]{ther:gram:2000}, and we give details for two methods
to detect it post-fit.

The first is to utilize \cite{whit:1980} who fitted the Cox model
as a Poisson GLM.  It relies on producing artificial data based on
the Poisson-multinomial `trick' (e.g., \cite{baker:1994}) so that
the Poisson and partial likelihoods are proportional to each other.
However, the setting up of many indicator variables results in
the need to estimate many nuisance parameters and the resulting
data set can be much larger than the original data set---all this
makes the procedure computationally expensive.  The supplementary
\RR{} script gives a numerical example of this method.

The second method is direct computation of the derivatives of the
observed information matrix
with respect to the~$\beta_k$
and using~(\ref{eq:deriv.matrix}).
This is quite manageable since the matrix has a simple form.
In the following we adopt a notation similar to \cite{lawl:2003}
and assume there are no ties or time-varying covariates for simplicity.
The data is of the form~$(y_i,\ \bix_i, \delta_i)$, $i=1,\ldots,n$,
containing~$k$ distinct
lifetimes~$y_{(1)} < \cdots < y_{(k)}$ and~$n-k$ censoring times.
Let~$R_i=R(y_{(i)})$ denote the
risk set at~$y_{(i)}$.
Let~$\delta_i=0$ or~1 for censored and complete survival times~$y_i$
respectively.
For individual~$i$ define $Z_i(y) = I(y_i \geq y)$ so
that~$Z_i(y) = 1$ if and only if~$i \in R(y)$.

The observed information matrix
is $\bA = \mbox{}$
\begin{eqnarray}
  \label{eq:lawl:2003:eqn.7.1.9.ver0}
  \sum_{i=1}^n \; \delta_i
  \left\{
\frac{
  \sum_{u =1}^n
  Z_u(y_i) \,
  \exp( \bbeta^T \bix_{u} )
  \left[ \bix_u - \overline{\bix}(y_i,\ \bbeta) \right]
  \left[ \bix_u - \overline{\bix}(y_i,\ \bbeta) \right]^T}{
  \sum_{u =1}^n
  Z_u(y_i) \,
  \exp( \bbeta^T \bix_{u} )}
\right\}  ~~~
\end{eqnarray}
where, for~$t>0$,
$
\overline{\bix}(t,\ \bbeta) = 
  \sum_{u=1}^n \; Z_u(y) \;
  \bix_{u} \,  \exp(\bbeta^T \bix_u)
/
   \sum_{u=1}^n \; Z_u(y) \, \exp(\bbeta^T \bix_u)
$.
Writing~$\bbeta=(\beta_1,\ldots,\beta_p)^T$
and~$\bix_i=(x_{i1},\ldots,x_{ip})^T$,
and letting the numerator of~(\ref{eq:lawl:2003:eqn.7.1.9.ver0})
be called~$\bcalA_i$, then
\begin{eqnarray}
\nonumber
  \frac{\partial \bA}{\partial \beta_k}
   &~=~&
  \sum_{i=1}^n \; \delta_i
  \left\{
\frac{
  \partial \bcalA_i / \partial \beta_k
}{
  \sum_{u =1}^n Z_u(y_i) \,   \exp( \bbeta^T \bix_{u} )
}
-
\frac{\bcalA_i  \;
\sum_{u =1}^n Z_u(y_i) \, x_{uk}  \exp( \bbeta^T \bix_{u} )
}{
\left[ \sum_{u =1}^n Z_u(y_i) \,   \exp( \bbeta^T \bix_{u} )\right]^2 }
\right\},
\\
\frac{\partial \bcalA_i }{ \partial \beta_k }
   &~=~&
\sum_{u =1}^n Z_u(y_i) \, x_{uk}  \exp( \bbeta^T \bix_{u} )
\left[ \bix_u - \overline{\bix}(y_i,\ \bbeta) \right]
\left[ \bix_u - \overline{\bix}(y_i,\ \bbeta) \right]^T
-
\mbox{} \nonumber \\ &&
\sum_{u =1}^n Z_u(y_i) \, \exp( \bbeta^T \bix_{u} ) \,
\left\{
\left[ \bix_u - \overline{\bix}(y_i,\ \bbeta) \right]
\frac{
  \partial \; \overline{\bix}(y_i,\ \bbeta)^T }{ \partial \beta_k }
+
\mbox{} \right. \nonumber \\ &&
\left.
  \frac{ \partial \; \overline{\bix}(y_i,\ \bbeta) }{ \partial \beta_k }
\left[ \bix_u - \overline{\bix}(y_i,\ \bbeta) \right]^T
\right\},
\nonumber
\\
\frac{ \partial \; \overline{\bix}(y_i,\ \bbeta) }{ \partial \beta_k }
&~=~&
\frac{
  \sum_{u =1}^n Z_u(y_i) \, x_{uk} \, \bix_u \, \exp( \bbeta^T \bix_{u} )
}{
  \sum_{u =1}^n Z_u(y_i) \, \exp( \bbeta^T \bix_{u} )
}
-
\mbox{} \nonumber \\ &&
\frac{
\left(
  \sum_{u =1}^n Z_u(y_i) \,           \bix_u \, \exp( \bbeta^T \bix_{u} )
\right)
\left(
  \sum_{u =1}^n Z_u(y_i) \, x_{uk} \,           \exp( \bbeta^T \bix_{u} )
\right)
}{
\left(
  \sum_{u =1}^n Z_u(y_i) \, \exp( \bbeta^T \bix_{u} )
\right)^2
} ~.
  \label{eq:coxph.oim.deriv1}
\end{eqnarray}

\subsection*{A5. Profile Likelihoods}

Suppose~$\btheta^T=(\btheta_1^T, \btheta_2^T)$
where~$\btheta_2$ comprise nuisance parameters,
and that~$\widehat{\btheta}_2 = \bgamma(\widehat{\btheta}_1)$
has continuous first derivatives.
Let~$\ell_c(\btheta_1) = \ell(\btheta_1,\bgamma({\btheta}_1))$
be the concentrated log-likelihood
and the observed information matrix be partitioned as
\begin{eqnarray}
\left(
\begin{array}{cc}
 \bA_{11} & \bA_{12} \\
 \bA_{21} & \bA_{22}
\end{array}
\right)^{-1} ~=~
\left(
\begin{array}{cc}
 \bA^{11} & \bA^{12} \\
 \bA^{21} & \bA^{22}
\end{array}
\right)^{}.
\label{eq:patterned.matrix.1}
\end{eqnarray}
Then the derivative of the Wald statistics
based on~$\ell_c$ requires
\begin{eqnarray}
\label{eq:conc.likelihood}
\frac{\partial \bA^{11}}{\partial \widehat{\beta}_s}
&~=~&
-\bA^{11} \,
\left[
\frac{\partial \bA_{11}}{\partial \widehat{\beta}_s} -
\frac{\partial \bA_{12}}{\partial \widehat{\beta}_s}\,
\bA_{22}^{-1} \,
\bA_{21} +
\mbox{} \right. \\ && \left.  
\bA_{12} \,
\bA_{22}^{-1} \,
\frac{\partial \bA_{22}}{\partial \widehat{\beta}_s} \,
\bA_{22}^{-1}
\bA_{21} -
\bA_{12} \,
\bA_{22}^{-1} \,
\frac{\partial \bA_{21}}{\partial \widehat{\beta}_s}
\right]
\bA^{11},
\nonumber
\end{eqnarray}
where~$\bA^{11} =
\left(\bA_{11}^{} - \bA_{12}^{} \, \bA_{22}^{-1}  \, \bA_{21}^{} \right)^{-1}$
and
$s \in \{1,\ldots,\dim(\btheta_1)\}$.
Some slight simplification follows by exploiting
symmetry through $\bA_{12} = \bA_{21}^T$, etc.

\subsection*{A6. Logistic Regression with a Binary Covariate}

It is shown here that it is possible to determine whether
the HDE will occur for a binary covariate
in a logistic regression.
The notation here differs slightly from the rest of the paper.
We wish to fit the logistic regression
\begin{eqnarray}
  \logit \; \pi ~=~
  \beta_1 \, x_1 + \bgamma^T \bix,
\label{eq:logistic.binary.x1}
\end{eqnarray}
where~$x_1=0$ or~1,
and~$\bix$  contains other covariates including the intercept.
The data can be summarized by Table~\ref{tab:Cgelimo:2x2table}, albeit
without being able to reflect the \bix.
Order the data so that~$i=1,\ldots,N_0$ for $x_1=0$ ,
and~$i=N_0+1,\ldots,N_0+N_1$ for $x_1=1$, and let
\begin{eqnarray*}
            {\pi}_{i0} &~=~& \expit\{           \bgamma^T \bix_i \}, \\
            {\pi}_{i1} &~=~& \expit\{ \beta_1 + \bgamma^T \bix_i \}.
\end{eqnarray*}
Then
\begin{eqnarray*}
\bX^T \bW \bX &~=~&
\bA ~=~
\left(
\begin{array}{cc}
a_{11} & \bia_{12} \\
\bia_{21} & \bA_{22}
\end{array}
\right)
  \\
  &~=~&
\left(
\begin{array}{cc}
\sum_{}^{}\,\widehat{\pi}_{i1}^{}(1-\widehat{\pi}_{i1})^{}&
\sum_{}^{}\,\widehat{\pi}_{i1}^{}(1-\widehat{\pi}_{i1})^{}\,\bix_i^T \\
\sum_{}^{}\,\widehat{\pi}_{i1}^{}(1-\widehat{\pi}_{i1})^{}\,\bix_i ~~~ &
\sum_{}^{}\,\widehat{\pi}_{i0}^{}(1-\widehat{\pi}_{i0})^{}\,\bix_i\,\bix_i^T +
\sum_{}^{}\,\widehat{\pi}_{i1}^{}(1-\widehat{\pi}_{i1})^{}\,\bix_i\,\bix_i^T
\end{array}
\right),
\end{eqnarray*}
where the summations are over the appropriate suffixes.
The (1, 1) element of its inverse is $a^{11} = \mbox{}$
\begin{eqnarray*}
\lefteqn{
\left[
\sum_{}^{} \, \widehat{\pi}_{i1}^{} (1-\widehat{\pi}_{i1})^{} -
\left(
\sum_{}^{}\,\widehat{\pi}_{i1}^{}(1-\widehat{\pi}_{i1})^{}\,\bix_i^T
\right)
\cdot
\right. } \\
&&
\left.
\left\{
\sum_{}^{}\,\widehat{\pi}_{i0}^{}(1-\widehat{\pi}_{i0})^{}\,\bix_i\,\bix_i^T +
\sum_{}^{}\,\widehat{\pi}_{i1}^{}(1-\widehat{\pi}_{i1})^{}\,\bix_i\,\bix_i^T
\right\}^{-1}
\left(
\sum_{}^{} \, \widehat{\pi}_{i1}^{} (1-\widehat{\pi}_{i1})^{} \, \bix_i
\right)
\right]^{-1}.
\end{eqnarray*}
Its first derivative with respect to $\beta_1$ is
\begin{eqnarray*}
(a^{11})'
  &~=~&
- (a^{11})^2 \cdot
\left\{
\sum_{}^{} \,
(1 - 2 \, \widehat{\pi}_{i1}^{})
\widehat{\pi}_{i1}^{} (1-\widehat{\pi}_{i1})^{} -
\mbox{} \right. \\ && \left.
\frac{\partial}{\partial \beta_1}
\left(
\sum_{}^{} \, \widehat{\pi}_{i1}^{} (1-\widehat{\pi}_{i1})^{} \, \bix_i^T
\right)
\bA_{22}^{-1}
\left(
\sum_{}^{} \, \widehat{\pi}_{i1}^{} (1-\widehat{\pi}_{i1})^{} \, \bix_i
\right)
\right\} \\
  &~=~&
- (a^{11})^2 \cdot
\calB,  \mbox{~~~say}.
\end{eqnarray*}
The last term is
\begin{eqnarray*}
\left(
\sum_{}^{} \,
(1 - 2 \, \widehat{\pi}_{i1}^{}) \,
\widehat{\pi}_{i1}^{} (1-\widehat{\pi}_{i1})^{} \, \bix_i^T
\right)
\bA_{22}^{-1}
\left(
\sum_{}^{} \, \widehat{\pi}_{i1}^{} (1-\widehat{\pi}_{i1})^{} \, \bix_i
\right)
+
\mbox{} && \\
\left(
\sum_{}^{} \,
\widehat{\pi}_{i1}^{} (1-\widehat{\pi}_{i1})^{} \, \bix_i^T
\right)
\frac{\partial \bA_{22}^{-1}}{\partial \beta_1}
\left(
\sum_{}^{} \, \widehat{\pi}_{i1}^{} (1-\widehat{\pi}_{i1})^{} \, \bix_i
\right)
+
\mbox{} && \\
\left(
\sum_{}^{} \,
\widehat{\pi}_{i1}^{} (1-\widehat{\pi}_{i1})^{} \, \bix_i^T
\right)
\bA_{22}^{-1}
\left(
\sum_{}^{} \,
(1 - 2 \, \widehat{\pi}_{i1}^{}) \,
\widehat{\pi}_{i1}^{} (1-\widehat{\pi}_{i1})^{} \, \bix_i
\right), &&
\end{eqnarray*}
where
\begin{eqnarray*}
\frac{\partial \bA_{22}^{-1}}{\partial \beta_1}
&~=~& -
\bA_{22}^{-1} \,
\frac{\partial \bA_{22}^{}}{\partial \beta_1} \,
\bA_{22}^{-1},
\\
\frac{\partial \bA_{22}^{}}{\partial \beta_1}
&~=~&
\sum_{}^{} \,
(1 - 2 \, \widehat{\pi}_{i1}^{}) \,
\widehat{\pi}_{i1}^{} (1-\widehat{\pi}_{i1})^{} \, \bix_i \, \bix_i^T.
\end{eqnarray*}
Thus the HDE will be present if
\begin{eqnarray}
\frac{\widehat{\beta}_1}{2}
\cdot a^{11} \cdot \calB
~<~  -1.
\label{eq:Cgelimo:hdeff.binary.xk}
\end{eqnarray}
The results of Section~\ref{sec:Cgelimo:hdeff.dispro} are
a special case of this.
In particular, suppose that~$\pi_0=1/2$ in
Table~\ref{tab:Cgelimo:hauc:donn:1977.dataset}(a) so that the
intercept in~(\ref{eq:Cgelimo:hauc:donn:1977.model}) vanishes and
need not be estimated.
Then straightforward calculations show that
the boundary where HDE occurs corresponds to the nonlinear
equation
$\logit \, {\widehat{\pi}_1}
  =
{2}/{(2 \, \widehat{\pi}_1 - 1)}
$.
Its numerical solution means that, approximately,
$|\widehat{\beta}_2| > 2.40$ is needed
in order for the HDE to occur.
For positive $\widehat{\beta}_2$,
this corresponds to an odds ratio of about~$11.0$.

\renewcommand{\arraystretch}{1.4}
\begin{table}[tt]
\caption{
A general $2\times 2$ table of counts
(there are actually covariates~\bix{} with each individual not
seen here.)
\label{tab:Cgelimo:2x2table}
}
\centering
 \begin{tabular}{lccr}
~~~ & $y=0$ & $y=1$ \\
\cline{2-3}
\multicolumn{1}{l|}{$x_{2}=0$~~} &
\multicolumn{1}{c|}{$N-R_0$} &
\multicolumn{1}{c|}{~~~$R_0$~~~}
& ~$N_0$
\\
\cline{2-3}
\multicolumn{1}{l|}{$x_{2}=1$~~} &
\multicolumn{1}{c|}{$N_1 - R_1$} &
\multicolumn{1}{c|}{~~~$R_1$~~~}
& ~$N_1$
   \\
\cline{2-3}
\end{tabular}
\end{table}
\renewcommand{\arraystretch}{1.0}

\bibliographystyle{plainnat}
 \bibliography{usebib}

\end{document}

%% file: pagelayoutWOmargins2.tex
\newcommand{\mk}[1]{\mbox{[1 mark]}}

\newcommand{\convergeD}{\mbox{$\stackrel{\cal D}{\longrightarrow}$}}

\newcommand{\calB}{\mbox{${\cal B}$}}

\newcommand{\calW}{\mbox{${\cal W}$}}

\newcommand{\stLM}{\mbox{\tiny  LM}}
\newcommand{\stVLM}{\mbox{\tiny VLM}}

\newcommand{\approxsim}{\mbox{$\stackrel{.}{\sim}$}}
\def\REAL{\hbox{\rm I \kern-5.5pt R}} 
\def\NAT{\hbox{\rm I \kern-5.5pt N}} 
\def\SREAL{\hbox{\rm I \kern-3.5pt R}} 
\def\SNAT{\hbox{\rm I \kern-3.5pt N}} 

\newcommand{\expit}{\mbox{\rm expit}}
\newcommand{\logit}{\mbox{\rm logit}}

\newcommand{\bone}{{\bf 1}}

\newcommand{\bia}{\mbox{$\bm{a}$}}

\newcommand{\bic}{\mbox{$\bm{c}$}}

\newcommand{\bie}{\mbox{$\bm{e}$}}

\newcommand{\bio}{\mbox{$\bm{o}$}}

\newcommand{\bix}{\mbox{$\bm{x}$}}

\newcommand{\biy}{\mbox{$\bm{y}$}}

\newcommand{\bA}{\mbox{\rm \bf A}}
\newcommand{\bB}{\mbox{\rm \bf B}}

\newcommand{\bH}{\mbox{\rm \bf H}}
\newcommand{\bI}{\mbox{\rm \bf I}}

\newcommand{\bL}{\mbox{\rm \bf L}}

\newcommand{\bO}{\mbox{\rm \bf O}}

\newcommand{\bR}{\mbox{\rm \bf R}}

\newcommand{\bU}{\mbox{\rm \bf U}}

\newcommand{\bX}{\mbox{\rm \bf X}}
\newcommand{\bW}{\mbox{\rm \bf W}}

\newcommand{\bbeta}{\mbox{\boldmath $\beta$}}

\newcommand{\bgamma}{\mbox{\boldmath $\gamma$}}
\newcommand{\bdelta}{\mbox{\boldmath $\delta$}}
\newcommand{\boldeta}{\mbox{\boldmath $\eta$}}

\newcommand{\bmu}{\mbox{\boldmath $\mu$}}

\newcommand{\btheta}{\mbox{\boldmath $\theta$}}

\newcommand{\bSigma}{\mbox{\boldmath $\Sigma$}}
\newcommand{\bTheta}{\mbox{\boldmath $\Theta$}}

\newcommand{\diag}{ \mbox{\rm diag} }
\newcommand{\Diag}{ \mbox{\rm Diag} }

\newcommand{\SE}{ \mbox{\rm SE} }

\newcommand{\sgn}{ \mbox{\rm sgn} }

\newcommand{\Cov}{ \mbox{\rm Cov} }
\newcommand{\Var}{ \mbox{\rm Var} }

\newcommand{\Corr}{ \mbox{\rm Corr} }

\newcommand{\pkg}[1]{\mbox{{\sf #1}}}

\newcommand{\RR}{{\textsf{R}}}

\def\Pr{{\rm Pr}}

\newcommand{\rcodett}[1]{{\texttt{#1}}}

\newcommand{\EIM}{\mbox{{\boldmath ${\cal I}$}${}_{E}$}}   

\newcommand{\EIMsub}[1]{\mbox{{\boldmath ${\cal I}$}${}_{E,#1}^{}$}}

\newcommand{\EIMone}{\mbox{{\boldmath ${\cal I}$}${}_{E1}$}}   

\newcommand{\EIMinvone}{\mbox{{\boldmath ${\cal I}$}${}_{E1}^{-1}$}}

\newcommand{\bcalA}{\mbox{{\boldmath ${\cal A}$}}}   %
